\documentclass[lettersize,journal]{IEEEtran}
\usepackage{amsmath,amsfonts}
\usepackage{algorithmic}
\usepackage{algorithm}
\usepackage{array}
\usepackage[caption=false,font=normalsize,labelfont=sf,textfont=sf]{subfig}
\usepackage{textcomp}
\usepackage{stfloats}
\usepackage{url}
\usepackage{verbatim}
\usepackage{graphicx}
\usepackage{cite}
\usepackage{enumitem}
\usepackage{placeins} 
\usepackage{amsmath}
\usepackage{amssymb} 
\usepackage{makecell}
\usepackage[hidelinks]{hyperref}
\usepackage{orcidlink}

\begin{document}
	
\title{GOMA: \underline{G}eometrically \underline{O}ptimal \underline{M}apping via \underline{A}nalytical Modeling for Spatial Accelerators}

\author{
	Wulve~Yang\,\orcidlink{0009-0003-1228-619X},
	Hailong~Zou\,\orcidlink{0009-0004-0090-9553},
	Rui~Zhou\,\orcidlink{0009-0001-8713-0005},
	Jionghao~Zhang\,\orcidlink{0009-0005-2139-0567},
	Qiang~Li\,\orcidlink{0009-0009-7082-6960},
	Gang~Li\,\orcidlink{0009-0004-8504-8149},
	Yi~Zhan\,\orcidlink{0009-0006-7304-3476}~\IEEEmembership{Member,~IEEE},
	and Shushan~Qiao\,\orcidlink{0000-0002-9102-2111}~\IEEEmembership{Member,~IEEE}%
	
	\thanks{Manuscript received xx; revised xx}%
	\thanks{This work was supported by the Beijing Natural Science Foundation under Grant L244052 and the National Science and Technology Major Project of Smart Grid under Grant 2025ZD0806400.\emph{(Corresponding authors: Yi Zhan; Shushan Qiao.)}}%
	\thanks{The authors are with the Institute of Microelectronics, Chinese Academy of Sciences, Beijing 100029, China, and also with the University of Chinese Academy of Sciences, Beijing 101408, China (e-mail: yizhan@ime.ac.cn; qiaoshushan@ime.ac.cn).}

}

	\markboth{IEEE TRANSACTIONS ON COMPUTERS, ~Vol.~x, No.~x, x~x}%
	{Shell \MakeLowercase{\textit{et al.}}: A Sample Article Using IEEEtran.cls for IEEE Journals}
	
	\IEEEpubid{0000--0000/00\$00.00~\copyright~2021 IEEE}
	\bstctlcite{mybstctl}	
	\maketitle
	\begin{abstract}
		General matrix multiplication (GEMM) on spatial accelerators is highly sensitive to mapping choices in both execution efficiency and energy consumption. However, the mapping space exhibits combinatorial explosion, which makes it extremely challenging to obtain optimal mappings within an acceptable time budget. Existing approaches  typically face challenges: They often lack global-optimality guarantees and become prohibitively slow as the mapping space grows.
		To address these limitations, we propose \textsc{GOMA}, a geometric-abstraction-based, globally optimal GEMM mapping framework via analytical modeling, which achieves efficient solving while guaranteeing optimality. \textsc{GOMA} introduces, from first principles, a geometric abstraction for GEMM mapping, yielding an exact analytical energy objective with $O(1)$ evaluation for any given mapping.
		The objective is highly accurate. \textsc{GOMA} then formulates mapping selection as an integer optimization problem under hardware and mapping constraints, using the analytical energy model as the objective to automate mapping search. \textsc{GOMA} can quickly compute a global-optimal mapping for any (GEMM workload, target hardware) pair, achieving this for the first time in mapping space exploration.
		Experiments confirm that across representative accelerators and large language model prefill workloads, \textsc{GOMA} improves the energy--delay product (EDP) by $2.24$--$4.24\times$ over SOTA mappers, while accelerating time-to-solution by $3.83$--$73.6\times$.

	\end{abstract}

	
	\section{Introduction}
	\label{sec:intro}
	
	\IEEEPARstart{W}{ith} the widespread adoption of deep learning, core operators such as general matrix multiplication (GEMM) and convolution have created enormous computational demand on both edge devices and data centers. Among them, GEMM dominates models such as Transformers and has become one of the most critical compute kernels. To meet this demand, spatial accelerators for DNN acceleration have become an important research direction in both academia and industry. This trend has led to classic designs such as Eyeriss~\cite{2016eyeriss}, Gemmini~\cite{2021gemmini}, and Google TPU~\cite{2017tpu}, as well as a large body of recent work~\cite{2025trex, 2025edge, 2025nebula, 2025mega}.
	
\begin{figure*}[!t]
		\centering
		\begin{minipage}[t]{0.49\textwidth}
			\centering
			\includegraphics[width=\linewidth]{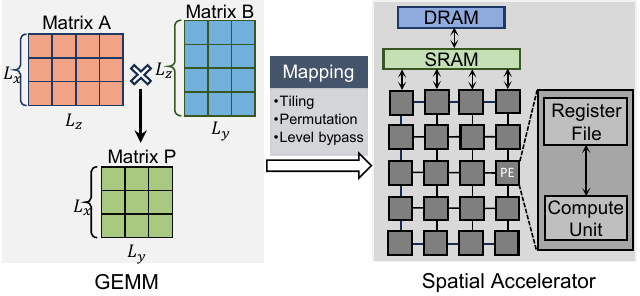}
			\caption{Mapping in a spatial accelerator. A mapping specifies tiling, loop permutation, and level bypass.}
			\label{fig:mapping-illustration}
		\end{minipage}
		\hfill
		\begin{minipage}[t]{0.49\textwidth}
			\centering
			\includegraphics[width=\linewidth]{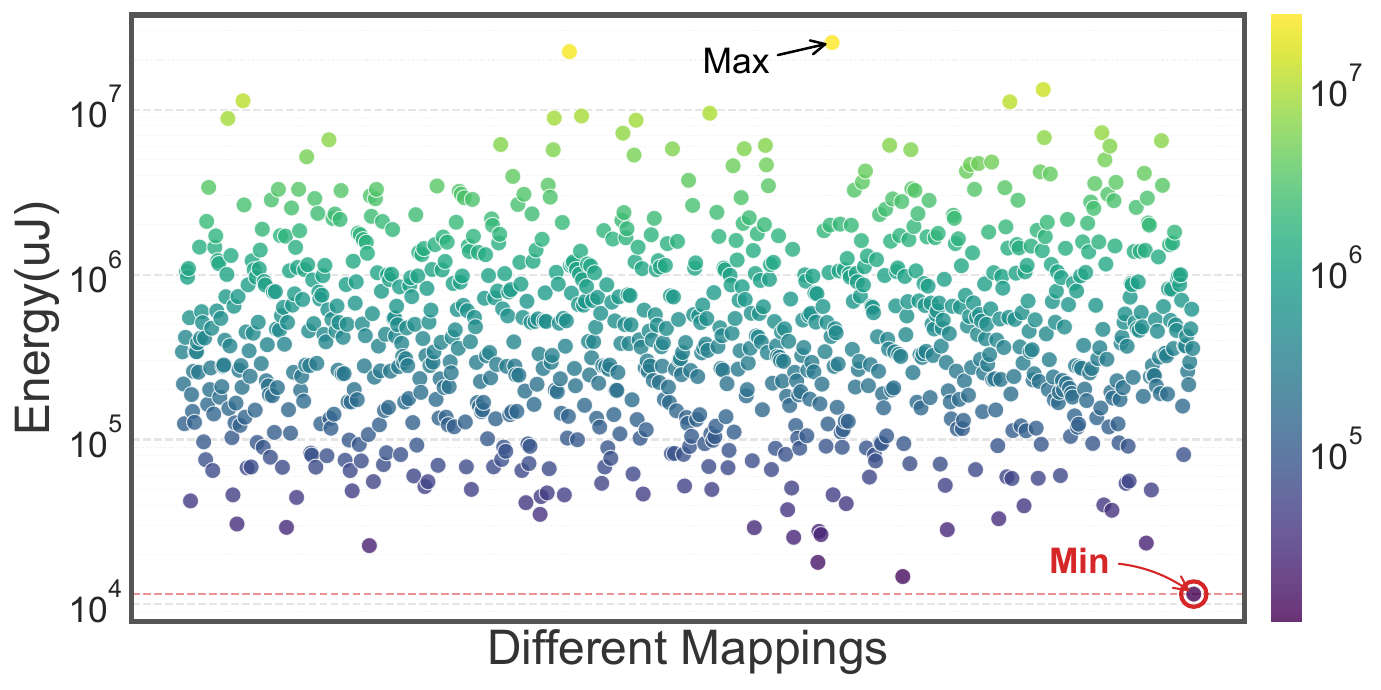}
			\caption{Energy variation across different mappings for the same GEMM on a spatial accelerator (log scale). Each point represents a mapping configuration.}
			\label{fig:performance-variation}
		\end{minipage}
	\end{figure*}

	In the hardware--algorithm co-design of such accelerators, \emph{mapping} is the key component that determines performance and energy efficiency. As illustrated in Fig.~\ref{fig:mapping-illustration}, a mapping describes the concrete execution strategy of a target algorithm on a specific hardware instance, including computation tiling, loop permutation, and level bypass.
	Mapping is both important and challenging. On the one hand, mapping choices can drastically affect the outcome. Even when deploying the same algorithm on the same hardware, different mappings can lead to differences of several orders of magnitude in energy efficiency and performance~\cite{2022digamma}. Fig.~\ref{fig:performance-variation} further shows that, for the same GEMM on the same spatial accelerator, mapping differences alone can induce orders-of-magnitude energy variation (for visualization convenience, the global extrema are omitted). On the other hand, the mapping space typically grows combinatorially. The mapping space of a typical convolution can exceed $10^{20}$~\cite{2022digamma}, while that of GEMM is far beyond $10^{10}$. Such scale makes exhaustive search for the optimal mapping computationally infeasible.
\IEEEpubidadjcol	

	Given the decisive impact of mapping on spatial accelerators, many mapping space exploration (MSE) methods have been proposed in recent years, including random search~\cite{2019timeloop,2019simba,2020interstellar}, black-box heuristic search~\cite{2021loma,2024tensormap,2023salsa}, differentiable model approximation~\cite{2021mindmappings,2023dosa}, and mathematical programming~\cite{2021cosa}. However, search-based methods often struggle to simultaneously achieve efficiency and solution quality in a massive discrete space. Differentiable approximations typically relax integer constraints, which introduces approximation errors and weakens result guarantees. Mathematical programming has the potential to enable exact solving in principle. But existing work (e.g., CoSA~\cite{2021cosa}) still has difficulty precisely characterizing real hardware costs, and the overall solving efficiency remains limited. Therefore, obtaining a provably optimal mapping within acceptable time remains an urgent open problem.
	
	To address the above challenges, we propose \textsc{GOMA}, a globally optimal GEMM mapping framework based on geometric abstraction and analytical modeling. This paper focuses on GEMM mapping for a widely used class of mainstream accelerator templates (Fig.~\ref{fig:mapping-illustration}), and aims to provide verifiable optimal solutions for this important application scenario. Since GEMM accounts for dominant computation in core workloads such as modern large language models (LLMs) and diffusion transformers (DiTs), \textsc{GOMA} is not only theoretically meaningful but also practically valuable. Through geometric abstraction and analytical modeling, \textsc{GOMA} precisely computes the cost of cross-level data movement and transforms the mapping problem into an integer optimization problem under hardware constraints. It thus enables fast global-optimal solving with a verifiable optimality certificate. This geometric representation is fundamentally different from prior approaches.

	Experimental results show that \textsc{GOMA} achieves solution quality and efficiency simultaneously. Across four representative accelerators and LLM prefill workloads, \textsc{GOMA} consistently achieves $2.24$--$4.24\times$ EDP improvements over multiple existing mapping methods.
	It also reduces solving time by $3.83$--$73.6\times$.
	As a result, it produces high-quality mappings with verifiable global optimality in seconds.
	
	Our main contributions are as follows:
	
	\begin{enumerate}
		\item We introduce a new compute-grid geometric abstraction for GEMM mapping, derived from first principles, which yields an exact analytical closed-form energy objective with $O(1)$ evaluation and 99.9\% consistency with timeloop-model~\cite{2019timeloop}.

		\item We propose \textsc{GOMA}, which formulates mapping selection uniformly as an integer optimization problem subject to hardware and mapping constraints, and uses the above energy model as the objective function to enable automated mapping search.
		
		\item For the first time in mapping optimization, \textsc{GOMA} can compute a global-optimal mapping for any given (GEMM workload, target hardware instance) pair within a short time.
		It also outputs a verifiable optimality certificate.
		Together, these results strictly guarantee global optimality under the modeled objective and constraints.
	\end{enumerate}

	We will open-source GOMA on GitHub after acceptance.

	\section{Related Works}
	
	To efficiently search a massive mapping space and locate high-quality mapping solutions, a large number of methods have been proposed in recent years. They can be broadly categorized as follows:
	
	\subsubsection{Random Search}
	These methods explore the space via random sampling or pruned random search. Their advantage is strong generality, as they can accommodate diverse hardware instances and constraints. However, due to limited sampling efficiency, they often fail to approach the optimum. Representative work includes Timeloop~\cite{2019timeloop}, Simba~\cite{2019simba}, and Interstellar~\cite{2020interstellar}.
	
	\subsubsection{Black-box Heuristic Search}
	This is one of the most widely used categories. It treats mapping parameters as inputs and performs iterative search using heuristic strategies coupled with a cost model. Existing work can be roughly divided into two groups:
	\emph{Single-Stage} methods directly iterate in the full space and update candidates (e.g., genetic algorithm (GA), reinforcement learning (RL), or feature-driven variants); representative studies include~\cite{2020gamma,2025senna,2025memorycentric,2021hasco,2023spotlight}.
	\emph{Multi-Stage} methods mitigate the efficiency issues of an excessively large space by introducing coarse-to-fine search, multi-stage pruning, or hybrid strategies; representative studies include~\cite{2024tensormap,2025factorflow,2023salsa,2025kapla}.
	Overall, in ultra-large discrete mapping spaces, heuristic methods typically provide no convergence guarantees, and their solutions often remain suboptimal. Moreover, they require repeated interactions with the cost model for evaluation, which can further lengthen the end-to-end search time.

	\subsubsection{Differentiable Model Approximation}
	To leverage efficient optimization algorithms such as gradient descent, some work (e.g.,~\cite{2021mindmappings, 2023dosa}) builds differentiable approximate cost models. However, this route typically faces two core challenges:
	\emph{(1) Approximation error}: the differentiable model deviates from the original discrete cost model.
	\emph{(2) Relaxation--rounding error}: integer variables, such as tiling factors, are first relaxed to continuous variables for solving, and the continuous solutions are then rounded/projected back to the integer domain, which breaks optimality. Moreover, mappings must satisfy strict divisibility constraints, so the rounded solutions require additional repairs to restore feasibility, which further degrade the results.

	\subsubsection{Pruned Enumeration}
	These methods can theoretically converge to the global optimum when search time is unconstrained. For example, LOMA~\cite{2021loma} and the linear-pruned strategy in timeloop-mapper~\cite{2019timeloop} reduce the candidate space by introducing pruning during traversal. However, such methods often require high search overhead and can be unaffordable for large networks or complex operator layers. To improve practicality, LOMA~\cite{2021loma} further proposes heuristic variants that trade part of optimality for usable search speed.

	\subsubsection{Mathematical Programming}
	These methods formulate mapping as a constrained optimization problem and hand it to a solver.
	If the objective faithfully reflects true hardware costs (e.g., energy), the formulation can, in principle, yield a globally optimal mapping.
	In practice, two limitations remain.
	First, the objective is often a proxy and can be misaligned with true energy.
	The root cause is that accurate cost modeling requires capturing the physical essence of mapping.
	This requirement is still not adequately addressed by existing methods.
	For example, CoSA~\cite{2021cosa} optimizes surrogate costs such as resource utilization rather than energy.
	Reproduction studies (e.g., FactorFlow~\cite{2025factorflow}) also report limited gains in many settings due to this misalignment.
	Second, the search-space representation can be redundant.
	CoSA uses prime-factor-level discrete encoding to cover the full scheduling space.
	However, it does not explicitly fold physically equivalent classes.
	A single physical execution may correspond to multiple equivalent encodings.
	As a result, solving still involves extensive combinatorial enumeration and repeated search.
	This redundancy slows down solving on large-scale problems.

	\emph{In addition,}
	some studies treat mapping space exploration as a component within a broader framework (e.g.,~\cite{2024multi,2024stream,2024feather,2024cocco}), and we do not elaborate on them here.

	Overall, for large-scale workloads, existing methods cannot obtain a globally optimal mapping within an acceptable time.

	\section{Intuition of \textsc{GOMA}}
	\label{sec:intuition}
	
	\begin{figure*}[t!]
		\centering
		\includegraphics[width=0.7\textwidth]{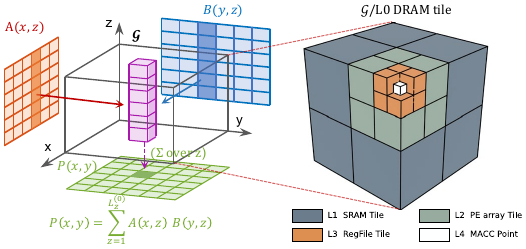}
		\caption{Geometric view of GEMM as a 3D compute grid and its three orthogonal projections corresponding to $A(x,z)$, $B(y,z)$, and partial sums/output $P(x,y)$. Mapping executes GEMM by hierarchically tiling the grid across the memory hierarchy.}
		\label{fig:gemm-geometry}
	\end{figure*}
	
	Because the formal definition of \textsc{GOMA} is notation-heavy, presenting it upfront is not conducive to intuitive understanding. Therefore, before presenting the formal definition, this section first introduces \textsc{GOMA}'s representation from an intuitive geometric perspective. It also explains why this representation captures the essence of mapping.
	
	By mapping/dataflow, we essentially refer to an execution strategy that can be implemented by hardware. It includes three key choices. \emph{Tiling} determines, at each level in the memory hierarchy, the amount of computation/data that resides at a time. This choice also implicitly specifies the spatial unrolling at the PE-array (processing-element array) level. \emph{Traversal order (loop permutation)} specifies how these tiles advance over time to cover the global compute space. \emph{Level bypass} decides whether a data type resides at a level to enable reuse, or bypasses this level and is supplied directly from an upper level to a lower level to avoid ineffective reads/writes. These dimensions are widely recognized as the core degrees of freedom in mapping space exploration (MSE)~\cite{2019timeloop,2025factorflow}. We develop a unified compute-grid geometric view, independently derived in this work, to systematically characterize these degrees of freedom.
	We then explain how this view determines access counts and energy. Based on it, \textsc{GOMA} jointly optimizes these choices and outputs a globally optimal solution with an independently verifiable certificate.

	\subsection{3D compute grid}
	Consider matrix multiplication:
	\begin{equation}
		P(x,y)=\sum_{z=1}^{L^{(0)}_z}A(x,z)\,B(y,z),\quad
		x\in[1,L^{(0)}_x],~y\in[1,L^{(0)}_y],
		\label{eq:mm}
	\end{equation}
	where the reduction happens along the $z$ axis. If we view one MAC (multiply--accumulate) as a compute-point $(x,y,z)$, then the entire computation can be represented as a global compute-point set $\mathcal{G}$, shown on the left of Fig.~\ref{fig:gemm-geometry}:
	\begin{equation}
		\mathcal{G}=\{(x,y,z)\mid x\in[1,L^{(0)}_x],~y\in[1,L^{(0)}_y],~z\in[1,L^{(0)}_z]\}.
		\label{eq:grid}
	\end{equation}
	The key value of this representation is as follows. In subsequent sections, ``how loops are nested / how parallelism is mapped'' can be uniformly transformed into ``how $\mathcal{G}$ is hierarchically covered and traversed.'' 
	
	\subsection{Data as Projections: Three Projections and Hierarchical Tiles}
	
	\subsubsection{Three Projections}
	As shown on the left of Fig.~\ref{fig:gemm-geometry}, each MAC in GEMM corresponds to one compute-point $(x,y,z)$ in the 3D compute grid. Under this geometric representation, the three matrices naturally align with three orthogonal projection planes. $A$ varies only with $(x,z)$, and can be viewed as the projection of the compute grid onto the $x$--$z$ plane. $B$ varies only with $(y,z)$, corresponding to the $y$--$z$ projection. The output/partial sums $P$ vary only with $(x,y)$, corresponding to the $x$--$y$ projection.
	For any 3D compute tile within $\mathcal{G}$, it covers a set of compute-points in $\mathcal{G}$ and thus corresponds to a sub-computation of GEMM. The same projection-alignment idea applies at the tile level. To compute one tile, we first obtain the data covered by this tile on the x--z and y--z planes, i.e., the two input projections. We then perform accumulation within the tile along the reduction axis z. The tile's result finally lands on the x--y plane, i.e., the P plane.
	Based on this correspondence, we use ``projection area'' to characterize the scale of data demand. For a 3D compute tile, its coverage on the three orthogonal planes ($x$--$z$, $y$--$z$, and $x$--$y$) gives the demand sizes of $A$, $B$, and $P$, respectively.
	
	\subsubsection{Hierarchical Tiles}
	Temporarily ignoring level bypass, a multi-level memory hierarchy can be understood as progressively covering the 3D compute grid $\mathcal{G}$ with hierarchical tiling, as shown on the right of Fig.~\ref{fig:gemm-geometry}. The hardware template used in this paper is shown in Fig.~\ref{fig:mapping-illustration}.
	It follows a five-level hierarchy: DRAM, SRAM, PE-array, regfile, and MACC (multiply--accumulate core).
	At the MACC level, a single MACC point $(x,y,z)$ is the minimal unit, corresponding to the smallest white cube in Fig.~\ref{fig:gemm-geometry}. In each step, the operands $A(x,z)$ and $B(y,z)$ are fed into the MACC, and accumulates the product into the partial sum $P(x,y)$. At the regfile level, the regfile of a single PE holds the $A/B/P$ required by a small compute tile. This allows the MACC in PE to continuously perform multiple MACs inside the tile, while registers keep supplying data, as shown by the orange cube. At the PE-array level, the PE-array spatially concatenates the small tiles of multiple PEs in parallel, forming a larger parallel compute tile, as shown by the green cube. At the SRAM level, SRAM holds larger-scale $A/B/P$ and repeatedly supplies data across multiple array phases, corresponding to the gray-blue cube. The outermost DRAM corresponds to the global compute-point set and stores the entire $A/B/P$ required by the matrix multiplication. Therefore, the overall execution can be summarized as follows: The global compute-point set (DRAM) is first partitioned into large SRAM tiles, then into PE-array tiles, then into regfile tiles, and finally MACs are performed point-by-point within each regfile tile.
	
	
	\subsubsection{Parallelism Stacking}
	Parallelism (\texttt{num\_pe}) constrains how PEs are spatially arranged along the $x/y/z$ axes (or their combinations), thereby shaping the array-level multicast sharing and spatial reduction patterns. 
	
	\subsection{How Traversal Determines Reuse: Walking Axis and Projection Update Counting}
	
\begin{figure}[t!]
		\centering
		\includegraphics[width=\columnwidth]{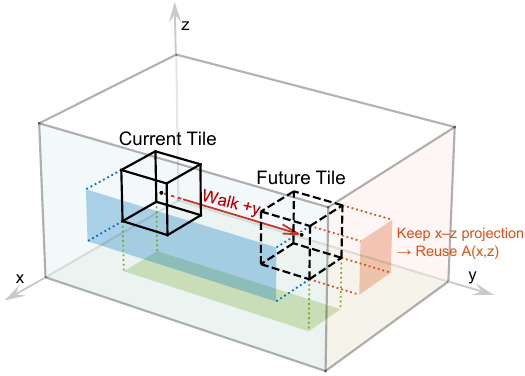}
		\caption{Walking-axis intuition. When a tile advances along one axis (here $y$), one projection (the $x$--$z$ plane) stays constant and can be reused, while the other two projections update.}
		\label{fig:walking-axis}
	\end{figure}

	Tiling answers only ``how much to compute each time,'' i.e., the local volume covered by one tile. To cover the entire 3D compute grid, we must also specify how 3D compute tiles advance in a higher level tile. The key observation is as follows. When a 3D compute tile advances by one step along an axis, exactly one of the three 2D projections stays unchanged, while the other two change. Combining this with the one-to-one correspondence between $A/B/P$ and the three projections, Fig.~\ref{fig:walking-axis} shows an example of advancing along the $y$ axis. In this case, the $x$--$z$ projection remains unchanged, which means $A(x,z)$ can achieve temporal reuse at this level. Similarly, when advancing along $x$, the $y$--$z$ projection stays unchanged, enabling continuous reuse of $B(y,z)$. When advancing along $z$, the $x$--$y$ projection stays unchanged, enabling reuse of $P(x,y)$ (partial sums).
	Therefore, why loop order affects energy can be reduced to the following. At a given level, which projection can remain unchanged for a longer time determines which data type can obtain stronger temporal reuse at that level. This reduces cross-level movement. We call ``the advancing direction that makes a projection remain unchanged during traversal at this level'' the walking axis of that level. Different levels may have different walking axes.
	
	From this viewpoint, when a tile moves from one position to the next, it usually does not need to re-transfer the whole tile. What truly needs to be transferred are the data slices newly replaced due to projection changes. Thus, total traffic naturally decomposes into three parts. We separately count the number of updates of the $y$--$z$ plane (corresponding to $B$), the $x$--$z$ plane (corresponding to $A$), and the $x$--$y$ plane (corresponding to $P$) during traversal. We then multiply these counts by the corresponding projection area to obtain the traffic volume. This counting further yields a stable pattern. The two projections parallel to the walking direction typically require frequent updates. In contrast, the projection orthogonal to the walking direction can often be compressed to ``once per column, only at the column head.'' The above assumes that the data reside at both this level and its adjacent upper level.
	
	\subsection{Bypass and the Reduction Axis: Path Rewriting and the Optimization Objective}
	
	\subsubsection{Bypass Jump Links}
	After introducing level bypass, the complexity is that it changes the actual layer-wise path taken by data. If a level chooses to bypass a data type, then this level no longer performs reads/writes or residency for that data type. The data is supplied directly from an upper level to a lower level, until it reaches the first level that chooses to store it. One direct impact is that the distribution of access counts across levels is rewritten. The bypassed level contributes zero accesses for that data type, but the load may shift to higher levels. Whether energy decreases depends on the trade-off between ``saving reads/writes at this level'' and ``increasing supply from upper levels.''
	
	More importantly, bypass makes the energy attribution of a transfer no longer confined to adjacent levels. For a given data type, its nearest upper resident level may be several levels away. Therefore, we no longer partition transfers or account energy using a fixed pair of adjacent levels. Instead, we determine the source–receiver pair per data type. This forms distinct cross-level hop links.
	Based on this, \textsc{GOMA}'s formal modeling is more naturally organized in a receiver-centric manner. We first answer ``how much data transfer is needed for this level's traversal.'' We then answer ``where each data item comes from.'' Different data sources lead to different per-transfer energy costs.
	
	\subsubsection{Specialness of the Reduction Axis}
	In matrix multiplication, the reduction happens along the $z$ axis. This makes the access semantics of the output/partial sums $P$ significantly different from those of the inputs $A,B$. For $A,B$, the typical behavior is reading from an upper level. For $P$, advancing along $z$ corresponds to an accumulation chain. On the same $(x,y)$, we repeatedly perform ``read the current partial sum $\rightarrow$ add the contribution of this MAC $\rightarrow$ write back the updated value.'' The first accumulation along $z$ does not need to read a previously computed partial sum. Therefore, we must distinguish boundary cases between the first step and subsequent steps. This makes counting and energy expansion more fine-grained. Detailed counting and energy expansion will be provided in later formal sections. Here we only emphasize the structural differences introduced by the reduction axis.
	
	\subsubsection{Optimization Problem}
	Combining the above intuition, \textsc{GOMA}'s optimization problem can be summarized as follows. Under hard constraints such as capacity, parallelism, and divisibility nesting, we choose (1) the tile shapes at each level (sizes along $x/y/z$), (2) the walking axis of each stage (which inner dimension advances), and (3) which data types reside or bypass at which levels, to minimize total energy. Energy can be computed in closed form due to a stable derivation chain. Computation corresponds to a 3D grid, and data corresponds to three projections. Traversal determines the update frequency of projections. Traffic is characterized by projection update counts and projection sizes. Energy is then a weighted result of traffic and unit costs. The core of \textsc{GOMA} is to reduce ``dataflow/mapping'' to a hierarchical geometric traversal problem, and to organize counting with projection updates as the unified object.
	
	\subsubsection{Remark}
	If extending to operators such as convolution, the ``compute grid'' has the potential to be generalized from 3D to higher dimensions. The above intuition still holds in essence.

	\section{GOMA Formulation}
	\label{sec:method}
	
\begin{figure*}[t!]
		\centering
		\includegraphics[width=\textwidth]{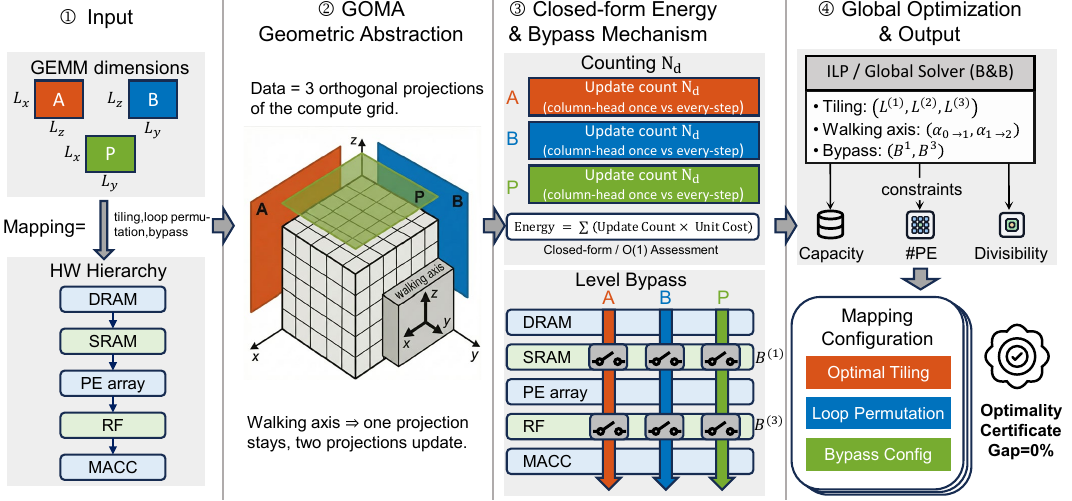}
		\caption{Overview of GOMA: geometric abstraction of mapping, closed-form update-count energy model with bypass gating, and global optimization that outputs the optimal tiling/dataflow/bypass together with an optimality certificate.}
		\label{fig:goma-overview}
	\end{figure*}

	This section formalizes the geometric representation and key observations from the previous chapter into a closed-form energy objective that can be used directly by a solver. These abstractions are developed from first principles in this work.
	Under hard constraints such as hardware capacity, parallelism, and divisibility/nesting, we jointly optimize hierarchical tiling, the walk axis (i.e., loop permutation), and per-axis bypass decisions to minimize the total energy. Concretely, we first abstract cross-level data movement as update counts of the three projections during traversal, and gate the relevant terms with bypass choices. We then weight these counts by hardware-provided per-word read/write energy costs and sum them up to obtain the total energy. Figure~\ref{fig:goma-overview} summarizes the overall workflow of \textsc{GOMA}.

	\subsection{Mapping Representation and Parameterization}
	
	\subsubsection{Notation: Compute Grid and Projections}
	\label{subsec:method-grid-proj}
	
	We follow the formalization of the GEMM compute domain from the previous section. The definition of matrix multiplication and the reduction axis is given in \eqref{eq:mm}, and the corresponding 3D compute grid is given in \eqref{eq:grid}.
	Based on the compute grid, $A(x,z)$, $B(y,z)$, and $P(x,y)$ can be viewed as projections onto three orthogonal coordinate planes.
	Specifically, $A\leftrightarrow x\text{--}z$, $B\leftrightarrow y\text{--}z$, and $P\leftrightarrow x\text{--}y$.
	
	To write the three data types uniformly using ``axis'' as the index in later expressions, we let
	\(
	d\in\{x,y,z\}
	\)
	index the normals of the three projection planes.
	The plane with normal \(x\) corresponds to the \(y\text{--}z\) projection, i.e., \(B\).
	The plane with normal \(y\) corresponds to the \(x\text{--}z\) projection, i.e., \(A\).
	The plane with normal \(z\) corresponds to the \(x\text{--}y\) projection, i.e., \(P\).
	Therefore, ``axis-wise counting/weighting'' is equivalent to ``matrix-wise counting/weighting.''
	
	\paragraph*{Units and Definition of Traffic Volume}
	\textsc{GOMA} uses a word as the minimum data granularity. In the following, ``traffic volume'' is uniformly measured in words: each projection element (i.e., one scalar of $A(x,z)$, $B(y,z)$, or $P(x,y)$) corresponds to one word.
	
	\subsubsection{Hardware Hierarchy and Hierarchical Tiling}
	\label{subsec:method-hierarchy}
	
	The hardware template used in this paper is shown in Fig.~\ref{fig:mapping-illustration}.
	We adopt a five-level abstraction from outer to inner.
	\begin{equation}
		\begin{aligned}
			p &\in \{0,1,2,3,4\} \\
			&\Rightarrow
			\{\text{DRAM},~\text{SRAM},~\text{PE-array},~\text{regfile},~\text{MACC}\}.
		\end{aligned}
		\label{eq:levels}
	\end{equation}
	The global compute-point size is
	\(
	(L^{(0)}_x,L^{(0)}_y,L^{(0)}_z)
	\).
	The lowest-level MACC corresponds to a single compute-point
	\(
	(L^{(4)}_x,L^{(4)}_y,L^{(4)}_z)=(1,1,1)
	\).
	At the SRAM/PE-array/regfile levels (\(p=1,2,3\)), we select tile lengths
	\(
	\{L^{(p)}_d\}_{d\in\{x,y,z\}}
	\)
	as tiling decision variables.
	
	Divisibility nesting between adjacent levels along each axis is captured by dimensionless ratios.
	\begin{equation}
		\hat L^{(p\!-\!p{+}1)}_{d}:=\frac{L^{(p)}_{d}}{L^{(p+1)}_{d}}\in\mathbb{N}_{+},
		d\in\{x,y,z\},~p\in\{0,1,2,3\}.
		\label{eq:hatL}
	\end{equation}
	We also use the global compute-point count (the total number of MACs), denoted as \(V\).
	\begin{equation}
		V:=L^{(0)}_x L^{(0)}_y L^{(0)}_z .
		\label{eq:V}
	\end{equation}
	
	\subsubsection{Stage-wise Walking Axes}
	\label{subsec:method-params}
	
	\paragraph{Stage-wise Walking Axis}
	As discussed earlier, the key impact of loop permutation can be characterized by which dimension a tile advances at a given level scale.
	\textsc{GOMA} assigns walking axes to stages to induce temporal traversal counts.
	\begin{equation}
		\alpha_{0-1}\in\{x,y,z\},\qquad
		\alpha_{1-2}\in\{x,y,z\},
		\label{eq:walking-axis}
	\end{equation}
	We use \(\{\beta,\gamma\}=\{x,y,z\}\setminus\{\alpha\}\) to denote the two axes orthogonal to the walking axis \(\alpha\); their order does not affect counting.
	
	\paragraph{Per-axis Bypass Policy}
	To capture whether a data type resides at a level for reuse or bypasses through it, we define a 2D binary switch matrix.
	\begin{equation}
		B \in \{0,1\}^{|\mathcal{D}|\times|\mathcal{P}|},
		\label{eq:bypass-matrix}
	\end{equation}
	Here, $B_{d,p}=1$ indicates that data with normal $d \in \mathcal{D}$ resides in memory level $p \in \mathcal{P}$.
	Bypass is applicable only to the SRAM and regfile levels; other levels do not support bypass.
	\begin{equation}
		B^{(0)}_{d}=B^{(2)}_{d}=B^{(4)}_{d}=1,\qquad
		B^{(1)}_{d},~B^{(3)}_{d}\in\{0,1\}.
		\label{eq:bypass-fixed}
	\end{equation}
	Intuitively, for each axis \(d\), the levels with \(B^{(p)}_{d}=1\) form the actual access chain for that axis.
	When a level is bypassed, its read/write and residency energy for that axis becomes zero.
	However, the axis data is supplied directly from a higher level to a lower level, which changes energy attribution.
	This motivates organizing energy accounting in a receiver-centric manner in the following.
	
	For brevity, \textsc{GOMA} uses an indicator function.
	\begin{equation}
		\mathbf{1}[\text{cond}]=
		\begin{cases}
			1,&\text{cond is true},\\
			0,&\text{otherwise}.
		\end{cases}
		\label{eq:indicator}
	\end{equation}
	
	\subsection{Closed-form Traffic Model}
	\label{subsec:method-counting}
	
	\subsubsection{Closed-form Update-Count Formulation}
	\textsc{GOMA} views cross-level communication as ``updates of projection data during traversal.''
	Under a walking axis \(\alpha\), the projection with normal \(\alpha\) can stay unchanged along the same ``column,'' so it can be compressed into ``count once at the column head.''
	The other two projections update more frequently when stepping.
	This mechanism directly corresponds to the geometric intuition from the previous section.
	
	\paragraph*{Note}
	For any 3D compute tile, the traffic (in words) of its input \(A/B\) equals the number of covered elements in the corresponding projection (i.e., the projection area).  
	For the output/partial sums \(P\), the total traffic of ``read old value + write back'' is proportional to \(P\)'s projection area, i.e., a constant multiple of that area.  
	This proportionality constant is uniformly accounted for by coefficients in the subsequent energy weights (e.g., \(\rho_z\)).

\subsubsection{src--1: DRAM \texorpdfstring{$\leftrightarrow$}{<->} SRAM}
\label{subsubsec:count-src1}

	The traffic volume of link $0$--$1$ along axis $d$ admits the following closed form:
	\begin{equation}
		\begin{aligned}
			N^{(0\!-\!1)}_{d}
			&=
			\mathbf{1}[B^{(1)}_{d}=1]\cdot V
			\\ &\quad\times
			\Big(
			L^{(0)}_{d}\cdot \mathbf{1}[d=\alpha_{0-1}]
			+
			L^{(1)}_{d}\cdot \mathbf{1}[d\neq \alpha_{0-1}]
			\Big)^{-1}
			\\ &\quad d\in\{x,y,z\}.
		\end{aligned}
		\label{eq:N01}
	\end{equation}
	
	Intuitively, when $d=\alpha_{0-1}$, the corresponding projection can be compressed to “once per global column head”.
	Hence the denominator is the global length $L^{(0)}_{d}$.
	Otherwise, the projection must be updated tile by tile in SRAM.
	Hence the denominator becomes the SRAM tile length $L^{(1)}_{d}$.
	The derivation of Eq.~\eqref{eq:N01} is deferred to the supplementary materials.

	\subsubsection{src--3: \texorpdfstring{$\cdots \leftrightarrow$}{... ->} Regfile}
	\label{subsubsec:count-src3}
	
	Only when regfile resides axis \(d\) (i.e., \(B^{(3)}_{d}=1\)) does there exist axis data transfer into regfile. Since \(2-3\) is spatial multicast from PE-array to regfile (which introduces no new temporal walking axis), its ``column-head compression'' is determined by the walking axis \(\alpha_{1-2}\) of the upper stage \(1-2\). The traffic volume is:
	\begin{equation}
		\begin{aligned}
			N^{(\mathrm{src}\!-\!3)}_{d}
			&=
			\mathbf{1}[B^{(3)}_{d}=1]\cdot
			\frac{V}{
				L^{(3)}_{d}\cdot
				\Big(\hat L^{(1\!-\!2)}_{d}\Big)^{\mathbf{1}[d=\alpha_{1-2}]}
			} \\
			&\quad d\in\{x,y,z\}.
		\end{aligned}
		\label{eq:Nsrc3}
	\end{equation}
	Here \(\hat L^{(1\!-\!2)}_{d}=L^{(1)}_{d}/L^{(2)}_{d}\).
	When \(d=\alpha_{1-2}\), the extra \(\hat L^{(1\!-\!2)}_{d}\) captures the compression effect.
	As the PE-array advances along this axis within an SRAM tile, regfile-side updates can be counted once per column head.
	The derivation of Eq.~\eqref{eq:Nsrc3} is deferred to the supplementary materials.
	
	\subsubsection{src--4: \texorpdfstring{$\cdots \leftrightarrow$}{... ->} MACC}
	\label{subsubsec:count-src4}
	
	At the MACC level, each global compute-point corresponds to one MAC. Therefore, for any axis \(d\), the number of MACC-side triggers is identical:
	\begin{equation}
		N^{(\mathrm{src}\!-\!4)}_{d}=V,\qquad d\in\{x,y,z\}.
		\label{eq:Nsrc4}
	\end{equation}
	The only difference is whether this supply comes from regfile, SRAM, or DRAM, which is determined by the per-axis bypass chain (see \S\ref{subsec:method-energy-aggregation}).
	
	\subsection{Boundary Handling for Reduction Axes}
	\label{subsec:method-boundary}
	
	The reduction axis \(z\) is special because updates of the output/partial sums \(P\) follow ``read old value, accumulate, and write back.''
	For a fixed \((x,y)\), if the receiving level experiences \(\tilde L^{(\mathrm{src}\!-\!p)}_{z}\) external updates, it writes back \(\tilde L^{(\mathrm{src}\!-\!p)}_{z}\) times.
	It reads the old value only \(\tilde L^{(\mathrm{src}\!-\!p)}_{z}-1\) times, since the first step can be treated as initialization from zero.
	
	\textsc{GOMA} defines the effective global column count \(\tilde L^{(\mathrm{src}\!-\!p)}_{z}\) for each receiver level \(p\in\{1,3,4\}\), where the dependence arises from walking axes and spatial sharing.
	\begin{equation}
		\tilde L^{(\mathrm{src}\!-\!1)}_{z}
		=
		\begin{cases}
			1, & \alpha_{0-1}=z,\\
			\dfrac{L^{(0)}_{z}}{L^{(1)}_{z}}, & \alpha_{0-1}\neq z,
		\end{cases}
		\label{eq:Ltilde-src1}
	\end{equation}
	\begin{equation}
		\tilde L^{(\mathrm{src}\!-\!3)}_{z}
		=
		\begin{cases}
			\dfrac{L^{(0)}_{z}}{L^{(1)}_{z}}, & \alpha_{1-2}=z,\\
			\dfrac{L^{(0)}_{z}}{L^{(2)}_{z}}, & \alpha_{1-2}\neq z,
		\end{cases}
		\label{eq:Ltilde-src3}
	\end{equation}
	\begin{equation}
		\tilde L^{(\mathrm{src}\!-\!4)}_{z}
		=
		\dfrac{L^{(0)}_{z}}{\hat L^{(2\!-\!3)}_{z}}.
		\label{eq:Ltilde-src4}
	\end{equation}
	
	For a fixed $(x,y)$, \(\tilde L\) represents the effective number of external updates required along the $z$ direction at receiver level $p$.
	When the walking axis is not $z$, an external interaction occurs once for each $z$-tile.
	Therefore, \(\tilde L\) equals the number of tiles along the global $z$ dimension
	(e.g., src--1 corresponds to $L^{(0)}_{z} / L^{(1)}_{z}$, and src--3 corresponds to $L^{(0)}_{z} / L^{(2)}_{z}$).
	When the receiver-level walking axis is $z$, partial sums along a $z$-column can be reused within the receiver level,
	which reduces \(\tilde L\).
	For src--4, the effective number of external updates is affected by spatial reduction,
	and thus manifests as a smaller effective column count.
	
	To avoid explicitly distinguishing ``first step'' from ``subsequent steps'' in the energy expression, we define a boundary coefficient.
	\begin{equation}
		\rho^{(\mathrm{src}\!-\!p)}_{z}
		:=
		\frac{\tilde L^{(\mathrm{src}\!-\!p)}_{z}-1}{\tilde L^{(\mathrm{src}\!-\!p)}_{z}}
		=
		1-\frac{1}{\tilde L^{(\mathrm{src}\!-\!p)}_{z}}.
		\label{eq:rho}
	\end{equation}
	
	The coefficient \(\rho\) is the ratio between the number of words read for the old value and the number of words written back at this receiver level.
	For a 3D compute tile, the projection area onto the plane with normal $z$ equals the write-back word count.
	With $\rho$, the read-old traffic can also be written as a scaled version of this projection area.
	Thus, read-old and write-back traffic are both counted using the same projection area on the plane with normal $z$.

	\subsection{Energy Weights from Hierarchical Access Costs}
	\label{subsec:method-energy-weights}
	
	Assume the hardware specification provides per-access read/write energy constants for each memory level.
	These include
	\(
	E^{\mathrm{DRAM}}_{\mathrm{read}},E^{\mathrm{DRAM}}_{\mathrm{write}},
	E^{\mathrm{SRAM}}_{\mathrm{read}},E^{\mathrm{SRAM}}_{\mathrm{write}},
	E^{\mathrm{regfile}}_{\mathrm{read}},E^{\mathrm{regfile}}_{\mathrm{write}}
	\).
	It also provides the MACC per-operation energy \(e^{\mathrm{MACC}}\).
	We encode the read/write energies triggered by one ``update with normal \(d\)'' across different levels into unit energy weights
	\(
	e^{(p,\downarrow)}_{d},e^{(p,\uparrow)}_{d}
	\)
	(the arrows indicate interactions with lower/higher levels).
	To match Timeloop's default accounting convention, when the lower-level memory writes back to the upper-level memory, we do not count the lower-level read energy.
	
	\subsubsection{0--1: DRAM \texorpdfstring{$\leftrightarrow$}{<->} SRAM} \mbox{}
	\label{subsubsec:weights-01}
	
	DRAM interacting with the lower level (layer \((0,\downarrow)\)):
	\begin{equation}
		\begin{aligned}
			e^{(0,\downarrow)}_{x} &= E^{\mathrm{DRAM}}_{\mathrm{read}},\\
			e^{(0,\downarrow)}_{y} &= E^{\mathrm{DRAM}}_{\mathrm{read}},\\
			e^{(0,\downarrow)}_{z} &= E^{\mathrm{DRAM}}_{\mathrm{write}}
			+ \rho^{(\mathrm{src}\!-\!p)}_{z}\cdot E^{\mathrm{DRAM}}_{\mathrm{read}}.
		\end{aligned}
		\label{eq:weights-dram-down}
	\end{equation}
	
	SRAM interacting with the upper level (layer \((1,\uparrow)\)):
	\begin{equation}
		\begin{aligned}
			e^{(1,\uparrow)}_{x} &= E^{\mathrm{SRAM}}_{\mathrm{write}},\\
			e^{(1,\uparrow)}_{y} &= E^{\mathrm{SRAM}}_{\mathrm{write}},\\
			e^{(1,\uparrow)}_{z} &= \rho^{(\mathrm{src}\!-\!p)}_{z}\cdot E^{\mathrm{SRAM}}_{\mathrm{write}}.
		\end{aligned}
		\label{eq:weights-sram-up}
	\end{equation}
	
	\subsubsection{1--2: SRAM \texorpdfstring{$\leftrightarrow$}{<->} PE-array} \mbox{}
	\label{subsubsec:weights-12}

	SRAM interacting with the lower level (layer \((1,\downarrow)\)):
	\begin{equation}
		\begin{aligned}
			e^{(1,\downarrow)}_{x} &= E^{\mathrm{SRAM}}_{\mathrm{read}},\\
			e^{(1,\downarrow)}_{y} &= E^{\mathrm{SRAM}}_{\mathrm{read}},\\
			e^{(1,\downarrow)}_{z} &= E^{\mathrm{SRAM}}_{\mathrm{write}}
			+ \rho^{(\mathrm{src}\!-\!p)}_{z}\cdot E^{\mathrm{SRAM}}_{\mathrm{read}}.
		\end{aligned}
		\label{eq:weights-sram-down}
	\end{equation}

	Following Timeloop's attribution (PE array as interconnect/fabric rather than storage), we do not count a separate PE-array-side access energy and set $e^{(2)}_{d}=0$.

	\begin{equation}
		e^{(2,\uparrow)}_{x}=e^{(2,\uparrow)}_{y}=e^{(2,\uparrow)}_{z}=0.
		\label{eq:weights-pe-up}
	\end{equation}
	
	\subsubsection{2--3: PE-array \texorpdfstring{$\leftrightarrow$}{<->} Regfile}\mbox{}
	\label{subsubsec:weights-23}
	
	Since $2\!\rightarrow\!3$ is spatial multicast, we likewise set the PE-array-side weights to zero:
	\begin{equation}
		e^{(2,\downarrow)}_{x}=e^{(2,\downarrow)}_{y}=e^{(2,\downarrow)}_{z}=0.
		\label{eq:weights-pe-down}
	\end{equation}
	
	Regfile interacting with the upper level (layer \((3,\uparrow)\)):
	\begin{equation}
		\begin{aligned}
			e^{(3,\uparrow)}_{x} &= E^{\mathrm{regfile}}_{\mathrm{write}},\\
			e^{(3,\uparrow)}_{y} &= E^{\mathrm{regfile}}_{\mathrm{write}},\\
			e^{(3,\uparrow)}_{z} &= \rho^{(\mathrm{src}\!-\!p)}_{z}\cdot E^{\mathrm{regfile}}_{\mathrm{write}}
			+E^{\mathrm{spa\_reduct}}.
		\end{aligned}
		\label{eq:weights-rf-up}
	\end{equation}
	Here \(E^{\mathrm{spa\_reduct}}\) denotes spatial reduction energy. Consistent with Timeloop's default implementation, we set \(E^{\mathrm{spa\_reduct}}=0\).
	
	\subsubsection{3--4: Regfile \texorpdfstring{$\leftrightarrow$}{<->} MACC} \mbox{}
	\label{subsubsec:weights-34}
	
	Regfile interacting with the lower level (layer \((3,\downarrow)\)):
	\begin{equation}
		\begin{aligned}
			e^{(3,\downarrow)}_{x} &= E^{\mathrm{regfile}}_{\mathrm{read}},\\
			e^{(3,\downarrow)}_{y} &= E^{\mathrm{regfile}}_{\mathrm{read}},\\
			e^{(3,\downarrow)}_{z} &= E^{\mathrm{regfile}}_{\mathrm{write}}
			+ \rho^{(\mathrm{src}\!-\!p)}_{z}\cdot E^{\mathrm{regfile}}_{\mathrm{read}}.
		\end{aligned}
		\label{eq:weights-rf-down}
	\end{equation}

	MACC is pure compute, so under \textsc{GOMA}'s accounting convention we do not count any storage-side access energy at level-4.
	Compute energy is accounted for in \S\ref{subsubsec:energy-macc}.

	\subsection{Receiver-centric Closed-form Energy Objective}
	\label{subsec:method-energy-aggregation}
	
	In this subsection, we multiply the update counts from \S\ref{subsec:method-counting} by the unit weights from \S\ref{subsec:method-energy-weights} and aggregate them.
	For optimization and comparison convenience, we define the normalized energy:
	\begin{equation}
		\bar E^{(\cdot)}:=\frac{E^{(\cdot)}}{V}.
		\label{eq:Enorm}
	\end{equation}
	
	\subsubsection{src--1 Term: \texorpdfstring{$\cdots \leftrightarrow$}{... ->} SRAM}
	\label{subsubsec:energy-src1}
	
	\begin{equation}
		\bar E^{(\mathrm{src}\!-\!1)}
		=
		\sum_{d\in\{x,y,z\}}
		\frac{N^{(0\!-\!1)}_{d}}{V}\cdot\Big(e^{(0,\downarrow)}_{d}+e^{(1,\uparrow)}_{d}\Big).
		\label{eq:Esrc1-norm}
	\end{equation}
	Here \(N^{(0\!-\!1)}_{d}\) is given by \eqref{eq:N01}.
	
	\subsubsection{src--3 Term: \texorpdfstring{$\cdots \leftrightarrow$}{... ->} Regfile (bypass-dependent source)}
	\label{subsubsec:energy-src3}
	
	When \(B^{(3)}_{d}=1\), the regfile must receive and hold the axis data. Its nearest upper resident level may be SRAM (if \(B^{(1)}_{d}=1\)) or DRAM (if \(B^{(1)}_{d}=0\)). Considering spatial multicast sharing in \(2-3\), the closed-form normalized energy is:
	\begin{equation}
		\begin{aligned}
			\bar E^{(\mathrm{src}\!-\!3)}
			&=
			\sum_{d\in\{x,y,z\}}
			\frac{N^{(\mathrm{src}\!-\!3)}_{d}}{V}\cdot
			\Big(
			e^{(3,\uparrow)}_{d}
			\\ &\quad
			+ \mathbf{1}[B^{(1)}_{d}=1]\cdot \frac{e^{(1,\downarrow)}_{d}}{\hat L^{(2\!-\!3)}_{d}}
			\\ &\quad
			+ \mathbf{1}[B^{(1)}_{d}=0]\cdot \frac{e^{(0,\downarrow)}_{d}}{\hat L^{(2\!-\!3)}_{d}}
			\Big).
		\end{aligned}
		\label{eq:Esrc3-norm}
	\end{equation}
	Here \(N^{(\mathrm{src}\!-\!3)}_{d}\) is given by \eqref{eq:Nsrc3}.
	
	\subsubsection{src--4 Term: \texorpdfstring{$\cdots \leftrightarrow$}{... ->} MACC (mutually exclusive sources)}
	\label{subsubsec:energy-src4}
	
	For each axis \(d\), MACC has only three mutually exclusive and collectively exhaustive direct sources: \\
	(i) regfile\(\leftrightarrow\)MACC: \(\mathbf{1}[B^{(3)}_{d}=1]\);\\
	(ii) SRAM\(\leftrightarrow\)MACC (via PE-array multicast): \(\mathbf{1}[B^{(1)}_{d}=1]\mathbf{1}[B^{(3)}_{d}=0]\);\\
	(iii) DRAM\(\leftrightarrow\)MACC (via PE-array multicast): \(\mathbf{1}[B^{(1)}_{d}=0]\mathbf{1}[B^{(3)}_{d}=0]\).
	Combining with \eqref{eq:Nsrc4}, the closed-form normalized energy is:
	\begin{equation}
		\begin{aligned}
			\bar E^{(\mathrm{src}\!-\!4)}
			&=
			\sum_{d\in\{x,y,z\}}
			\Big(
			\mathbf{1}[B^{(3)}_{d}=1]\cdot e^{(3,\downarrow)}_{d}
			\\ &\quad
			+ \mathbf{1}[B^{(1)}_{d}=1]\mathbf{1}[B^{(3)}_{d}=0]\cdot
			\frac{e^{(1,\downarrow)}_{d}}{\hat L^{(2\!-\!3)}_{d}}
			\\ &\quad
			+ \mathbf{1}[B^{(1)}_{d}=0]\mathbf{1}[B^{(3)}_{d}=0]\cdot
			\frac{e^{(0,\downarrow)}_{d}}{\hat L^{(2\!-\!3)}_{d}}
			\Big).
		\end{aligned}
		\label{eq:Esrc4-norm}
	\end{equation}
	
	\subsubsection{Compute Energy Term}
	\label{subsubsec:energy-macc}
	
	Each compute-point triggers one MAC. Therefore:
	\begin{equation}
		\bar E^{(4)}=e^{\mathrm{MACC}}.
		\label{eq:Emacc}
	\end{equation}
	
	\subsubsection{Leakage Energy Term}
	\label{subsubsec:energy-leak}
	
	Let \(\text{num\_pe}\) denote the number of PEs, satisfying:
	\begin{equation}
		\hat L^{(2\!-\!3)}_x\hat L^{(2\!-\!3)}_y\hat L^{(2\!-\!3)}_z=\text{num\_pe}.
		\label{eq:numpe}
	\end{equation}
	Let the per-cycle leakage constants be \(e^{\mathrm{SRAM}}_{\mathrm{leak}}\) and \(e^{\mathrm{RF}}_{\mathrm{leak}}\). Then the normalized leakage energy is:
	\begin{equation}
		\bar E^{(\mathrm{leak})}
		=
		\frac{
			e^{\mathrm{SRAM}}_{\mathrm{leak}}
			+
			e^{\mathrm{RF}}_{\mathrm{leak}}\cdot \text{num\_pe}
		}{\text{num\_pe}}.
		\label{eq:leak-total}
	\end{equation}
	For a fixed hardware instance and fixed \(\text{num\_pe}\), \(\bar E^{(\mathrm{leak})}\) is a constant and usually does not affect the optimal mapping.

	\subsection{Optimization Formulation}
	
	\subsubsection{Constraints} \mbox{}
	\label{subsec:method-constraints}
	
	\paragraph{Capacity Constraints (bypassed data excluded)}\mbox{}
	
	Regfile (level-3) capacity:
	\begin{equation}
		C^{(3)} \ge
		B^{(3)}_{y}L^{(3)}_xL^{(3)}_z
		+
		B^{(3)}_{x}L^{(3)}_yL^{(3)}_z
		+
		B^{(3)}_{z}L^{(3)}_xL^{(3)}_y .
		\label{eq:cap-rf}
	\end{equation}
	
	SRAM (level-1) capacity:
	\begin{equation}
		C^{(1)} \ge
		B^{(1)}_{y}L^{(1)}_xL^{(1)}_z
		+
		B^{(1)}_{x}L^{(1)}_yL^{(1)}_z
		+
		B^{(1)}_{z}L^{(1)}_xL^{(1)}_y .
		\label{eq:cap-sram}
	\end{equation}
	
	\paragraph{PE Number Constraints}
	Given by \eqref{eq:numpe}.
	
	\paragraph{Integrality and Divisibility Constraints}
	For any axis \(d\) and adjacent levels \(p\in\{0,1,2,3\}\), we require \(\hat L^{(p\!-\!p{+}1)}_d\in\mathbb{N}_+\) in \eqref{eq:hatL}.
	Equivalently, \(L^{(p+1)}_d\) must divide \(L^{(p)}_d\).
	
	\subsubsection{Problem Statement and Complexity}
	\label{subsec:method-opt}
	
	Combining \eqref{eq:Esrc1-norm}--\eqref{eq:Esrc4-norm} and \eqref{eq:Emacc} (and optionally \eqref{eq:leak-total}), we define the total normalized energy.
	\begin{equation}
		\bar E_{\text{total}}
		=
		\bar E^{(\mathrm{src}\!-\!1)}
		+
		\bar E^{(\mathrm{src}\!-\!3)}
		+
		\bar E^{(\mathrm{src}\!-\!4)}
		+
		\bar E^{(4)}
		\;(+\bar E^{(\mathrm{leak})}).
		\label{eq:Etotal}
	\end{equation}
	We then write mapping search as an optimization problem that minimizes this objective under constraints.
	
	\begin{equation}
		\begin{aligned}
			\min_{\substack{
					L^{(1)},L^{(2)},L^{(3)}\in\mathbb{N}_+^{3},\\
					\alpha_{0-1},\alpha_{1-2}\in\{x,y,z\},\\
					B^{(1)},B^{(3)}\in\{0,1\}^{3}
			}}
			\quad &
			\bar E_{\text{total}}
			\\
			\text{s.t.}\quad &
			\text{capacity constraints \eqref{eq:cap-rf}--\eqref{eq:cap-sram}}
			\\
			&
			\text{PE number constraints \eqref{eq:numpe}}
			\\
			&
			\text{divisibility constraints \eqref{eq:hatL}.}
		\end{aligned}
		\label{eq:opt-prob}
	\end{equation}

	A key advantage is that for any fixed decision
	\(\{L^{(p)}_d,\alpha_{0-1},\alpha_{1-2},B^{(1)},B^{(3)}\}\),
	energy evaluation reduces to a finite set of substitutions and summations over \(d\in\{x,y,z\}\).
	With a fixed number of hierarchy levels and data types, its computation complexity is constant (\(O(1)\)) and does not grow with the problem size or the number of tiles.
	Therefore, \textsc{GOMA} can serve as a fast scoring function for mapping space exploration, consistent with the geometric intuition from the previous section (``projection update counts \(\times\) energy weights'').
	
	\subsection{Optimality Justification}
	\label{sec:optimality-justification}
	
	We have formulated MSE as a constrained energy minimization problem.
	To claim the obtained solution is an optimal mapping in the modeling sense, two conditions must be satisfied.
	\begin{enumerate}
		\item {High fidelity of the objective and constraints.}\\
		The closed-form energy evaluator and constraints should reflect the reference energy model and the feasible mapping space as faithfully as possible.
		This ensures the solved problem aligns with the intended semantics.
		
		\item {Verifiable solver-level global optimality.}\\
		The solving process should provide a verifiable proof of optimality, so that the returned solution is indeed globally optimal under the objective and constraints.
		
	\end{enumerate}
	
	When these two conditions hold, we obtain an optimal mapping for the formulated MSE problem.
	
	\subsubsection{Fidelity of the Objective}
	We use Timeloop's energy model as a reference implementation (proxy oracle) to evaluate the fidelity of the closed-form energy objective. Timeloop reports that its analytical model is highly consistent with real measurements~\cite{2019timeloop}. Therefore, using timeloop-model as a reference baseline for hardware energy modeling is reasonable. Our goal is to verify the consistency between \textsc{GOMA}'s closed-form energy model and timeloop-model under identical settings.
	
	To cover energy variation induced by different GEMM shapes and mapping degrees of freedom, we design an evaluation set that spans tiling allocation, walking-axis selection, and per-level storage bypass policies.
	We select seven representative matrix multiplication operators from Llama-3.2-1B(1k) and map them onto an Eyeriss-like accelerator.
	For each GEMM, we construct 1152 ``tiling--permutation (walking axis)--bypass'' combinations.
	In total, we obtain 8064 mapping configurations.
	For each configuration, under the same energy reference table (ERT) and the same mapping semantics, we compute the total energy using both the closed-form evaluator and timeloop-model. We then evaluate consistency from both the pointwise error distribution and the overall energy-weighted error.
	
	Using the relative error $\frac{|E^{\textsc{GOMA}}-E^{\text{TL}}|}{E^{\text{TL}}}$ as the metric, among the 8064 mappings, 8004 (99.26\%) are exactly consistent with timeloop-model (relative error equals 0). Across all samples, the mean relative error is 0.099\% (corresponding to an energy consistency of $\sim$99.9\%).
	The median as well as the $p95/p99$ percentiles are all 0, indicating that errors are highly sparse and overall consistency is extremely high.
	 Furthermore, we report the energy-weighted overall relative error:
	$\sum_i |E^{\textsc{GOMA}}_i-E^{\text{TL}}_i| \big/ \sum_i E^{\text{TL}}_i$,
	which is 0.066\%. These results indicate that, under settings covering broad mapping degrees of freedom, \textsc{GOMA}'s closed-form energy objective can reproduce timeloop-model's energy evaluation with near pointwise consistency.
	
	\subsubsection{Solver-level Global Optimality}
	Under constraints such as capacity, parallelism, and divisibility, we formulate mapping search as a constrained optimization problem with integer variables, and solve it using Gurobi with branch-and-bound.
	For each instance, the solver maintains and outputs the objective upper bound (upper bound) corresponding to the current best feasible solution, the provable objective lower bound (lower bound), and the optimality gap between them.
	We terminate only when the gap converges to $0$. This yields directly verifiable proof information (UB/LB and gap), ensuring that the reported optimal mapping is globally optimal under the modeled objective and constraints.

	\section{Evaluation}
	\label{sec:eval}
	
	This section evaluates the end-to-end effectiveness of \textsc{GOMA} on realistic and representative GEMM workloads in the LLM prefill phase and on multiple spatial accelerator templates.
	We focus on the energy--delay product (EDP) and compare against multiple state of the art (SOTA) and prior-SOTA mapping space exploration (MSE) methods.
	
	\subsection{Experimental Setup}
	\label{subsec:eval-setup}
	
	\subsubsection{Workloads}
	\label{subsubsec:eval-workloads}
	
	\textsc{GOMA} targets GEMM as the core operator.
	We select representative large models covering both edge and center/cloud deployments and use the prefill phase as the evaluation scenario.
	
	\begin{itemize}[leftmargin=1.2em]
		\item \textbf{Edge LLMs:} Qwen3-0.6B and LLaMA-3.2-1B.
		\item \textbf{Center LLMs:} Qwen3-32B and LLaMA-3.3-70B.
	\end{itemize}
	
	The input sequence length significantly changes the \(X/Y/Z\) sizes of GEMMs, and thus changes the optimal mapping.
	Therefore, for each model category, we evaluate three input-length settings (short/medium/long).
	For edge scenarios, we use \(\{1\text{k}, 8\text{k}, 32\text{k}\}\).
	For center scenarios, we use \(\{2\text{k}, 32\text{k}, 128\text{k}\}\).
	This yields 12 workloads in total.
	
	For each workload, we enumerate all matrix multiplication operators involved in the prefill phase.
	We then group them into eight types:
	\texttt{attn\_q\_proj}, \texttt{attn\_kv\_proj}, \texttt{attn\_score}, \texttt{attn\_context}, \texttt{attn\_output},
	\texttt{mlp\_gate\_up}, \texttt{mlp\_down}, and \texttt{lm\_head}.
	We treat each type of GEMM as one mapping instance, and run each mapper to obtain its EDP.
	
	It is important to note that Transformer blocks with the same shape are stacked multiple times in the model, and a GEMM of the same shape may also appear 1--2 times within a single block.
	Therefore, the case-level EDP we report is an occurrence-count-weighted aggregation rather than a simple average over the eight operator types.
	
	\begin{equation}
		\mathrm{EDP}_{\text{case}} \triangleq \sum_{g \in \mathcal{T}} w_g \cdot \mathrm{EDP}(g),
	\end{equation}
	Here $\mathcal{T}$ denotes the eight GEMM types, and $w_g$ denotes the occurrence count of type $g$ in the prefill computation graph.
	We derive $w_g$ from model structural parameters and source-code parsing (e.g., \#layers and \#heads).
	The resulting counts are consistent with statistics from the actual computation graph.
	
	By default, we instantiate GEMMs using 8-bit quantized weights and activations.
	
	\subsubsection{Target Accelerators}
	\label{subsubsec:eval-arch}
	
	We select four representative spatial accelerator templates and model them under the unified \texttt{timeloop/accelergy} framework:
	two edge-oriented templates (Eyeriss-like and Gemmini-like) and two center-oriented templates (A100-like and TPU v1-like).
	For A100-like, we abstract L1/L2 cache hierarchy as global buffer (GLB), and apply equivalent scaling only to the array compute corresponding to Tensor Cores.
	Table~\ref{tab:eval-arch} summarizes the key resource and technology parameters.

	\begin{table}[h]
		\centering
		\caption{Key parameters of evaluated accelerator templates.}
		\label{tab:eval-arch}
		\setlength{\tabcolsep}{3pt}
		\renewcommand{\arraystretch}{1.2} 
		\setcellgapes{3pt}
		\makegapedcells
		\begin{tabular}{lccccc}
			\hline
			\textbf{Accelerator}
			& \textbf{\makecell{GLB\\(KiB)}}
			& \textbf{\makecell{Spatial\\Fanout\\(\#PE)}}
			& \textbf{\makecell{RF\\(words/PE)}}
			& \textbf{\makecell{Tech\\(nm)}}
			& \textbf{DRAM} \\
			\hline
			Eyeriss-like & 162   & 256   & 424 & 65 & LPDDR4 \\
			Gemmini-like & 576   & 256   & 1   & 22 & LPDDR4 \\
			A100-like    & 36864 & 65536 & 128 & 7  & HBM2   \\
			TPU v1-like  & 30720 & 65536 & 2   & 28 & DDR3  \\
			\hline
		\end{tabular}
	\end{table}
	
	To reflect platform deployment differences, we construct 24 cases.
	The six edge workloads are evaluated only on the two edge templates (\(6\times 2=12\)).
	The six center workloads are evaluated only on the two center templates (\(6\times 2=12\)).
	This yields 24 cases in total.
	
	\subsubsection{Baselines}
	\label{subsubsec:eval-baselines}
	
	We compare against the following representative mapping algorithms/frameworks:
	\textbf{Timeloop-mapper (Hybrid)}, \textbf{LOMA}~\cite{2021loma}, \textbf{SALSA}~\cite{2023salsa}, \textbf{CoSA}~\cite{2021cosa}, and \textbf{FactorFlow}~\cite{2025factorflow}.
	For each method, we prioritize the authors' recommended settings or the default parameters of public implementations. For SALSA, since its default configuration fails to converge within a long time in center-side experiments, we moderately reduce its configuration to ensure convergence while minimizing impact on mapping quality.
	
	It is important to emphasize that most baseline methods do not explicitly search per-matrix residency/bypass (level bypass) decisions.
	Therefore, when running these baselines, we enforce the bypass constraints specified by hardware. In contrast, \textsc{GOMA} and Timeloop Hybrid support modeling and searching bypass decisions, and thus we do not impose additional preset restrictions.
	
	\begin{figure*}[t!]
	\centering
	\includegraphics[width=\textwidth]{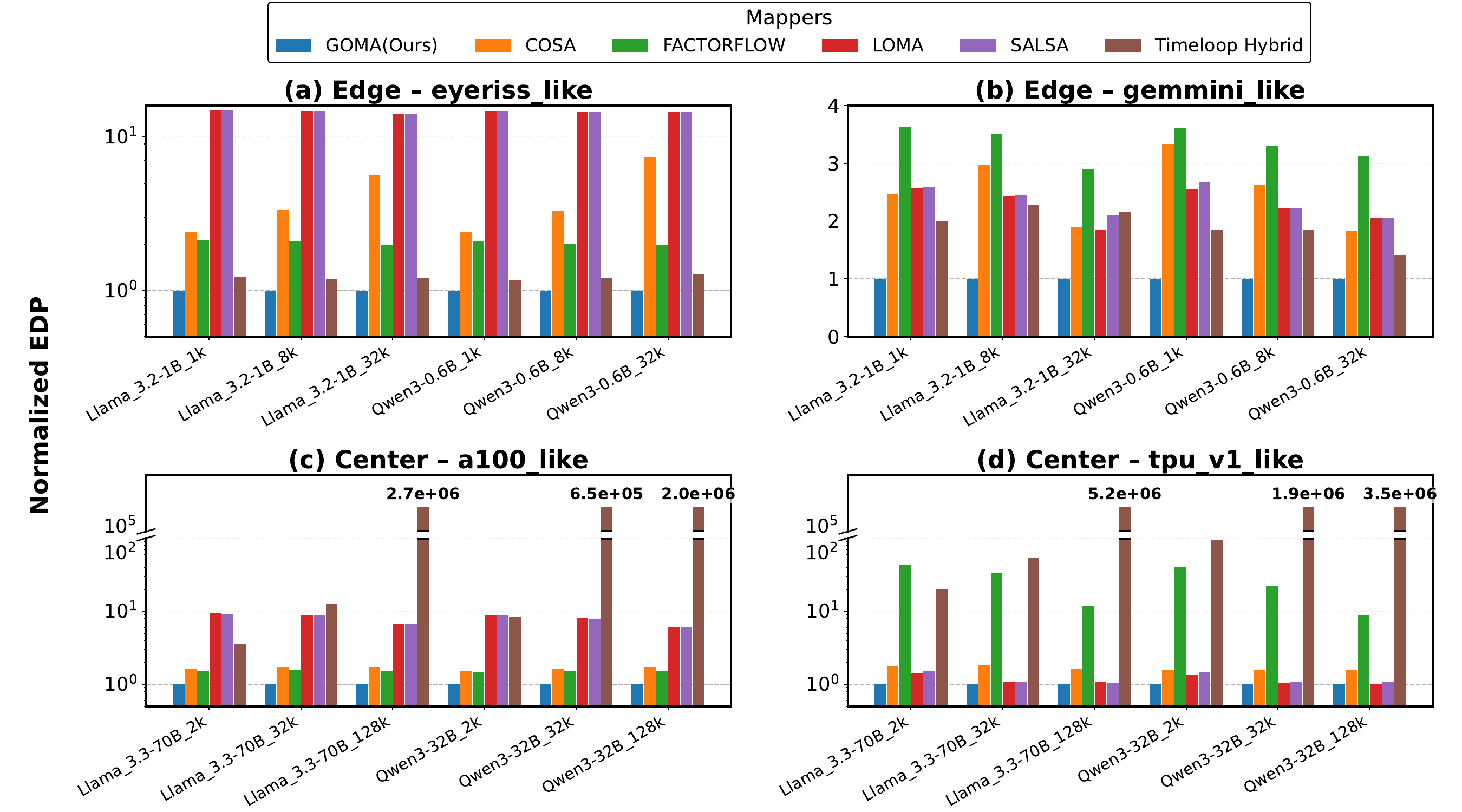}
	\caption{Overall normalized EDP comparison across 12 LLM prefill workloads on four accelerator templates. All bars are normalized to GOMA (lower is better).}
	\label{fig:eval-edp-overall}
	\end{figure*}
	
	\subsubsection{Evaluation Flow and Metrics}
	\label{subsubsec:eval-flow}

	To ensure consistent evaluation, we source all energy parameters from the Accelergy-generated energy reference table (ERT).
	This includes per-access memory energy, compute energy, and leakage energy.

	We use the energy--delay product (EDP) as the overall metric.
	\begin{equation}
		\mathrm{EDP} = E \times T.
	\end{equation}
	Although our optimization objective is the total energy $E$, we report $EDP=E\times T$ for evaluation to align with prior work.
	Under the PE-number constraint in Eq.~\eqref{eq:numpe}, the mappings found by \textsc{GOMA} achieve $100\%$ PE utilization, so $T$ reaches the compute lower bound.
	Therefore, under this setting, minimizing $E$ is equivalent to minimizing $EDP$, and the two yield the same ranking and conclusions.
	
	For aggregated visualization across cases, we report normalized EDP.
	\begin{equation}
		\mathrm{EDP}_{\mathrm{norm}}(\cdot) = \frac{\mathrm{EDP}(\cdot)}{\mathrm{EDP}(\mathrm{GOMA})},
	\end{equation}
	Under this normalization, \textsc{GOMA} is always 1 in all figures.
	Larger values indicate worse EDP relative to \textsc{GOMA}.
	
	For fair comparison, we use timeloop-model as a unified oracle to report 
	$E$, $T$, and EDP for both GOMA and all baselines. When reporting the final runtime, we do not include the additional verification time of timeloop-model.
	\paragraph*{Experimental Environment}
	All experiments are conducted on a laptop equipped with an AMD Ryzen 7 7840H processor. The reported runtime is the wall-clock time for running mapping search on this device.
	For fairness, we use a Python environment without free-threading enabled (GIL preserved).
	
	\begin{figure*}[t]
	\centering
	\includegraphics[width=\textwidth]{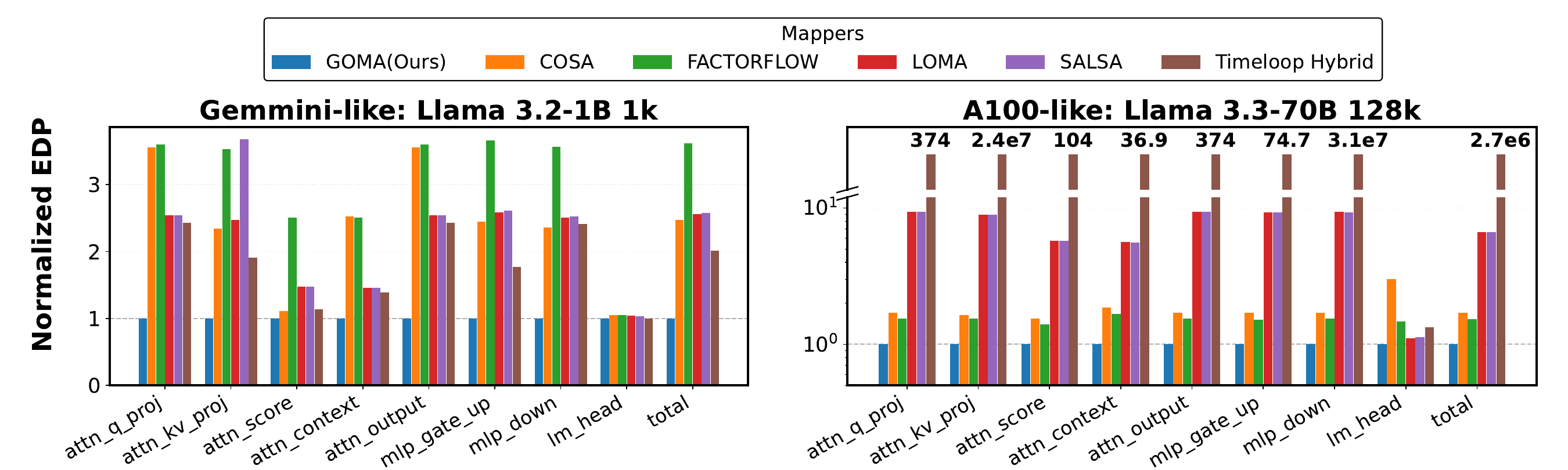}
	\caption{Per-layer (per-GEMM) normalized EDP breakdown for two representative cases. Bars are normalized to GOMA within each layer.}
	\label{fig:eval-edp-layer}
	\end{figure*}
	\subsection{EDP Results}
	\label{subsec:eval-edp}

	\subsubsection{Overall Comparison}
	\label{subsubsec:eval-edp-overall}
	
	Fig.~\ref{fig:eval-edp-overall} shows the normalized EDP comparison across four accelerator templates and 24 cases.

	From the overall results, we draw three key conclusions.
	
	\paragraph{Consistent wins across the full landscape}
	\textsc{GOMA} achieves the lowest EDP for all workloads and all accelerator templates, showing robust advantages across model scales (0.6B$\rightarrow$70B).
	The same robustness holds across sequence lengths (1k$\rightarrow$128k) and hardware resources (256$\rightarrow$65k PEs), despite large GLB/RF variations.
	Table~\ref{tab:eval-edp-summary} summarizes the geometric mean (geomean) and median over 24 cases.
	In terms of geomean, the closest baseline to \textsc{GOMA} is CoSA, yet it still lags by \(2.24\times\).
	Other baselines exhibit gaps of roughly \(3.9\times\sim 4.2\times\).
	
	\begin{table}[h]
		\centering
		\captionsetup{justification=centering}
		\caption{Summary of normalized EDP over 24 evaluation cases (normalized to GOMA; lower is better).}
		\label{tab:eval-edp-summary}
		\setlength{\tabcolsep}{3pt} 
		\renewcommand{\arraystretch}{1.2} 
		\setcellgapes{3pt}
		\makegapedcells
		\begin{tabular}{lcccccc}
			\hline
			\textbf{Metric} & \textbf{GOMA} & \textbf{CoSA} & \textbf{FactorFlow} & \textbf{LOMA} & \textbf{SALSA} & \textbf{\makecell{Timeloop\\Hybrid}} \\
			\hline
			Geomean & 1.00 & 2.24 & 3.91 & 4.17 & 4.24 & 98.5 \\
			Median  & 1.00 & 1.83 & 2.51 & 4.31 & 4.37 & 2.95 \\
			\hline
		\end{tabular}
	\end{table}

	\paragraph{Heuristic methods exhibit significant instability}
	Heuristic and multi-stage search methods such as LOMA, SALSA, and FactorFlow can approach the optimum in some cases, but deviate substantially in others, showing workload-dependent fluctuations.
	This is consistent with the structure of our analytical objective in \S\ref{sec:method}: the objective is jointly influenced by multi-level tiling, loop permutation, and discrete divisibility/capacity constraints, forming a highly non-convex and discontinuous combinatorial landscape. Purely relying on local neighborhood search and empirical pruning is difficult to stably approach the global optimum.
	
	\paragraph{Bypass is a key degree of freedom affecting EDP}
	Timeloop Hybrid achieves the best overall performance besides \textsc{GOMA} on Eyeriss-like and Gemmini-like templates.
	This largely benefits from its ability to explore combinations of level bypass across multiple memory levels during search, which reduces unnecessary residency and yields additional energy and performance gains.
	
	\paragraph{Remark}
	On large-scale arrays and deeper memory hierarchies such as A100-like and TPU v1-like, Timeloop Hybrid becomes noticeably unstable in some cases. The normalized EDP can reach up to the \(10^6\) level.
	This behavior mainly arises because Timeloop Hybrid's search mechanism has difficulty converging when the mapping space expands sharply.
	
	\subsubsection{Per-layer Breakdown}
	\label{subsubsec:eval-edp-layer}
	
	Each case contains multiple GEMM instances, so the overall bar chart does not clearly show which operator shapes contribute most to the gap.
	Fig.~\ref{fig:eval-edp-layer} selects two representative cases to present per-layer (per-GEMM) EDP comparisons.
	The first is Gemmini-like + LLaMA-3.2-1B (1k), representing smaller edge workloads.
	The second is A100-like + LLaMA-3.3-70B (128k), representing ultra-large center workloads.

	\paragraph{}From the per-layer breakdown, we observe that \texttt{lm\_head} has a ``matrix--vector'' shape.
	This makes it easier for multiple mappers to approach the optimum.
	In the Gemmini-like example, the gap between different methods and \textsc{GOMA} on \texttt{lm\_head} is small.
	Timeloop Hybrid can even achieve the optimum, which is consistent with the fact that such GEMMs have lower effective dimensionality and fewer feasible tilings, and thus are more favorable for pruning/heuristic methods.

	\paragraph{}Matrix--matrix GEMMs are the main source of gaps, and the advantage amplifies with scale.
	In the Gemmini-like example, \textsc{GOMA} already shows stable leads over baselines on multiple matrix--matrix workloads such as \texttt{attn\_q\_proj}, \texttt{attn\_kv\_proj}, \texttt{mlp\_gate\_up}, and \texttt{mlp\_down}.
	In larger-scale settings such as A100-like + 70B/128k, these leads are further amplified on major large GEMMs.
	This indicates that, as scale increases, combinatorial explosion in the mapping space makes sampling/heuristic methods harder to reliably find high-quality solutions, while analytical modeling plus global solving provides stronger determinism and guarantees.

	\subsection{Mapper Runtime}
	\label{subsec:eval-runtime}
	
	\subsubsection{Overall Runtime}
	\label{subsubsec:eval-runtime-overall}
	
\begin{figure*}[t]
		\centering
		\includegraphics[width=\textwidth]{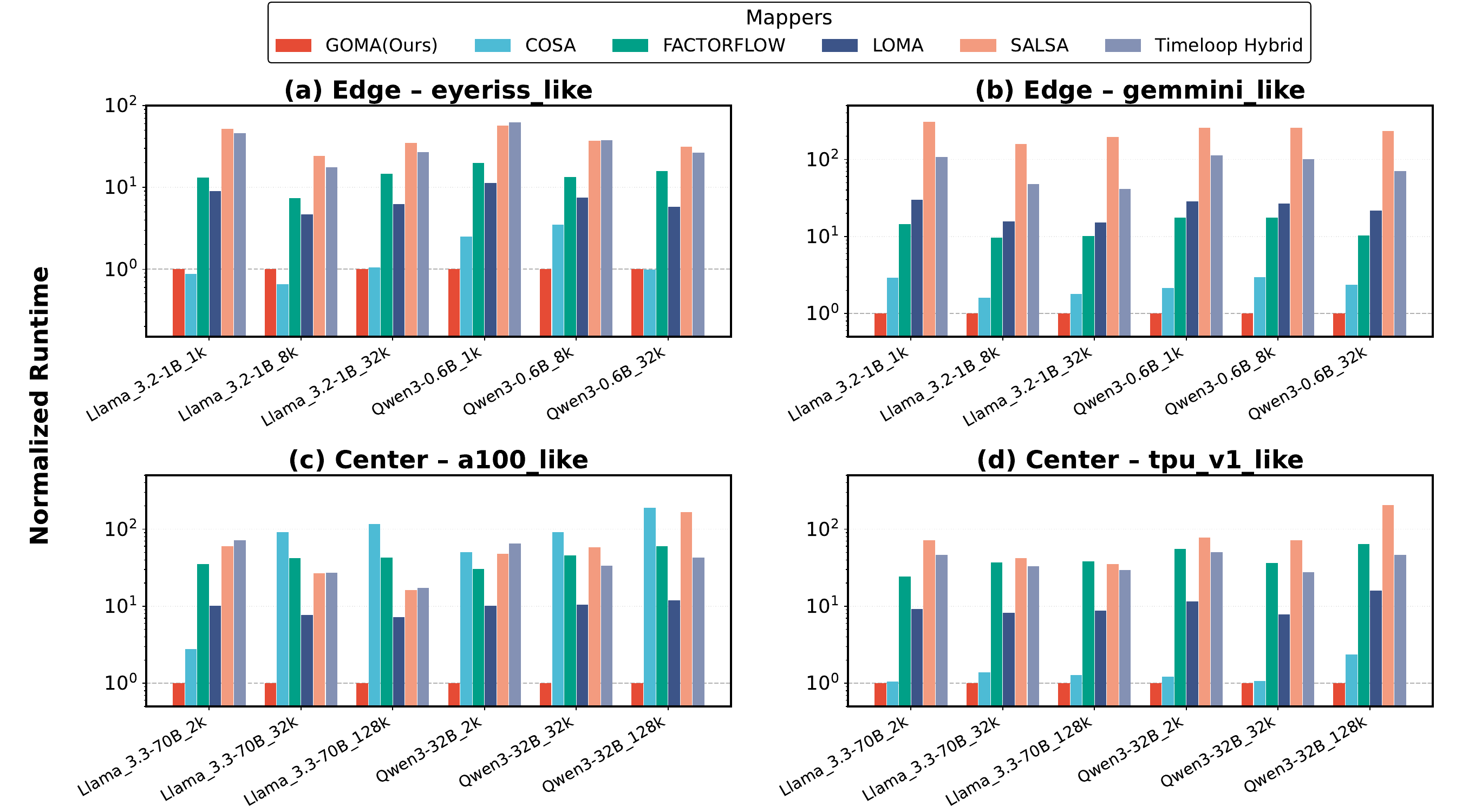}
		\caption{Overall normalized mapper runtime across 24 evaluation cases. All bars are normalized to GOMA (lower is faster).}
		\label{fig:eval-runtime-overall}
	\end{figure*}

	Beyond mapping quality, solving/search speed is also a core metric for deploying automated mapping.
	Fig.~\ref{fig:eval-runtime-overall} presents the normalized runtime of each mapper across 24 cases (lower is better; all normalized to \textsc{GOMA}).
	
	Table~\ref{tab:eval-runtime-summary} summarizes the geomean runtime.
	Overall, in our experimental setting, \textsc{GOMA} not only finds the optimal solution but is also the fastest mapper, showing a clear time advantage.
	Across the 24 cases (each case contains 8 GEMMs), the case-level runtime of \textsc{GOMA} has a geomean of 5.22\,s.
	This averages to only 0.65\,s per GEMM, and the maximum per-layer runtime is only 3.6\,s, meeting the needs of real-time mapping.

	\begin{table}[h]
		\centering
		\captionsetup{justification=centering}
		\caption{Summary of normalized mapper runtime over 24 evaluation cases}
		\label{tab:eval-runtime-summary}
		\setlength{\tabcolsep}{3pt} 
		\renewcommand{\arraystretch}{1.2} 
		\setcellgapes{3pt}
		\makegapedcells
		\begin{tabular}{lcccccc}
			\hline
			\textbf{Metric} & \textbf{GOMA} & \textbf{CoSA} & \textbf{FactorFlow} & \textbf{LOMA} & \textbf{SALSA} & \textbf{\makecell{Timeloop\\Hybrid}} \\
			\hline
			Geomean & 1.00 & 3.83 & 23.3 & 11.0 & 73.6 & 43.5 \\
			\hline
		\end{tabular}
	\end{table}

	From Fig.~\ref{fig:eval-runtime-overall}, we also observe that, on smaller edge cases, CoSA's runtime can sometimes be close to \textsc{GOMA},
	and it may even slightly outperform on a few points. However, as workload scale and resource scale increase, CoSA and other baselines generally exhibit substantial runtime blow-up.
	In contrast, \textsc{GOMA}'s analytical objective has constant-time evaluation (\S\ref{sec:method}), and it uses explicitly folded low-dimensional integer decision variables.
	This enables more efficient branch-and-bound pruning, maintaining stable and controllable solving time even for large-scale problems.
	
	\subsubsection{Case Study--GOMA vs.\ CoSA on Large Workloads}
	\label{subsubsec:eval-runtime-case}
	
\begin{figure*}[t]
		\centering
		\includegraphics[width=\textwidth]{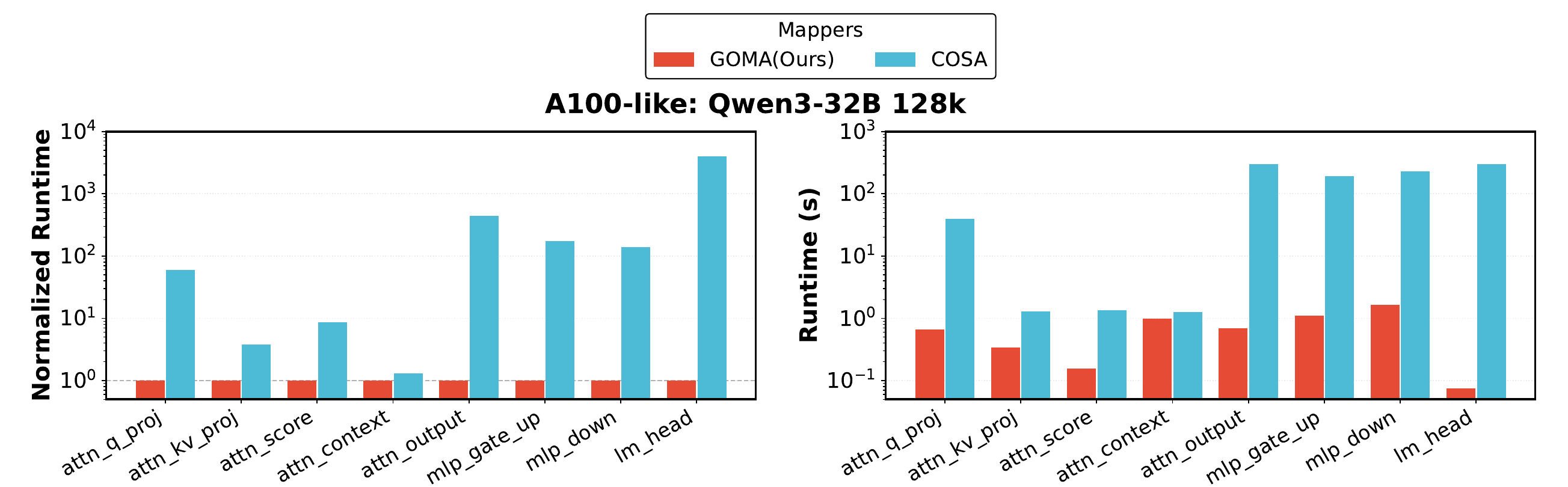}
		\caption{Runtime comparison between GOMA and CoSA on A100-like: Qwen3-32B(128k). Left: normalized runtime (to GOMA). Right: absolute runtime in seconds.}
		\label{fig:eval-runtime-layer}
	\end{figure*}

	CoSA is the modeling-based baseline with the closest geomean runtime to \textsc{GOMA}, and it belongs to the same broad family of methods.
	To highlight scale effects, we further select a case such as A100-like + Qwen3-32B (128k).
	Fig.~\ref{fig:eval-runtime-layer} compares the per-layer runtime across the eight GEMMs in this case (left: runtime normalized to \textsc{GOMA}; right: absolute seconds).
	To avoid excessively long runtimes in a few layers harming comparability, we set a per-layer time limit of 300\,s for CoSA in this case.
	Even so, for multiple large GEMMs (e.g., \texttt{attn\_output}, \texttt{mlp\_gate\_up}, \texttt{mlp\_down}, and \texttt{lm\_head}), CoSA still reaches the hundreds-of-seconds range.
	
	This comparison reflects the essential difference between the two modeling strategies.
	CoSA's prime-factor-level discrete encoding resembles implicit combinatorial enumeration, which often causes solving time to grow sharply with scale for high-dimensional GEMMs.
	In contrast, \textsc{GOMA} directly models at the geometric tiling scale and expresses energy costs in closed-form analytical expressions.
	As a result, the optimization variable dimension depends mainly on the fixed number of hierarchy levels.
	It is only weakly related to the numeric scales of \(X/Y/Z\).
	This enables stable and very short solving time even for ultra-large workloads.

\section{Conclusion}
We propose \textsc{GOMA}, a unified framework for GEMM mapping optimization on spatial accelerators.
It spans mapping specification, energy evaluation, and globally optimal solving.
Specifically, \textsc{GOMA} uses geometric abstraction to reduce hierarchical execution to closed-form counting of projection updates.
It also rewrites cross-level access chains via per-axis residency/bypass (level bypass).
This enables $O(1)$ analytical energy evaluation for any given mapping.
Building on this, \textsc{GOMA} jointly formulates tiling, loop permutation, and bypass as an integer optimization problem subject to hard constraints such as capacity, parallelism, and divisibility nesting.
It then outputs a verifiable optimality certificate via global solving, thereby strictly guaranteeing global optimality under the modeled objective and constraints.
	Experimental results show that \textsc{GOMA}'s analytical model is highly consistent with timeloop-model. Across multiple representative hardware templates and LLM prefill workloads, \textsc{GOMA} consistently achieves the lowest EDP and delivers $2.24$--$4.24\times$ EDP improvements and $3.83$--$73.6\times$ time-to-solution speedups over existing SOTA mappers.
	
As deployment demand for spatial accelerators continues to grow, \textsc{GOMA}'s analytical modeling and globally optimal solving capabilities provide a reusable foundation for more complex scenarios.
These scenarios include multi-layer mapping exploration and software--hardware co-optimization search.
	
	\bibliographystyle{IEEEtran}  
	\bibliography{refs}           

@IEEEtranBSTCTL{mybstctl,
	CTLuse_forced_etal       = "yes",
	CTLmax_names_forced_etal = "6",
	CTLnames_show_etal       = "6"
}

@inproceedings{2016eyeriss,
	title={{Eyeriss}: A Spatial Architecture for Energy-Efficient Dataflow for Convolutional Neural Networks ({CNN}s)},
	author={Chen, Yu-Hsin and Krishna, Tushar and Emer, Joel S. and Sze, Vivienne},
	booktitle={IEEE/ACM International Symposium on Computer Architecture (ISCA)},
	pages={367--379},
	year={2016}
}

@inproceedings{2021gemmini,
	title={{Gemmini}: Enabling Systematic Deep-Learning Architecture Evaluation via Full-Stack Integration},
	author={Genc, Hasan and Kim, Seah and Amid, Alon and Haj-Ali, Ameer and Iyer, Vighnesh and Prakash, Pranav and Zhao, Jerry and Grubb, Daniel and Liew, Harrison and Mao, Howard and Ou, Albert and Schmidt, Colin and Steffl, Samuel and Wright, John and Stoica, Ion and Ragan-Kelley, Jonathan and Asanovi{\'c}, Krste and Nikoli{\'c}, Borivoje and Shao, Yakun Sophia},
	booktitle={ACM/IEEE Design Automation Conference (DAC)},
	pages={769--774},
	year={2021}
}

@inproceedings{2017tpu,
	title={In-Datacenter Performance Analysis of a {Tensor Processing Unit}},
	author={Jouppi, Norman P. and Young, Cliff and Patil, Nishant and Patterson, David and Agrawal, Gaurav and Bajwa, Raminder and Bates, Sarah and Bhatia, Suresh and Boden, Nan and Borchers, Al and Boyle, Rick and Cantin, Pierre-Luc and Chao, Clifford and Clark, Chris and Coriell, Jeremy and Daley, Mike and Dau, Matt and Dean, Jeffrey and Gelb, Ben and Ghaemmaghami, Tara V. and Gottipati, Rajendra and Gulland, William and Hagmann, Robert and Ho, C. Richard and Hogberg, Doug and Hu, John and Hundt, Robert and Hurt, Dan and Ibarz, Julian and Jaffey, Aaron and Jaworski, Alek and Kaplan, Alexander and Khaitan, Harshit and Koch, Andy and Kumar, Naveen and Lacy, Steve and Laudon, James and Law, James and Le, Diemthu and Leary, Chris and Liu, Zhuyuan and Lucke, Kyle and Lundin, Alan and MacKean, Gordon and Maggiore, Adriana and Mahony, Maire and Miller, Kieran and Nagarajan, Rahul and Narayanaswami, Ravi and Ni, Ray and Nix, Kathy and Norrie, Thomas and Omernick, Mark and Penukonda, Narayana and Phelps, Andy and Ross, Jonathan and Ross, Matt and Salek, Amir and Samadiani, Emad and Severn, Chris and Sizikov, Gregory and Snelham, Matthew and Souter, Jed and Steinberg, Dan and Swing, Andy and Tan, Mercedes and Thorson, Gregory and Tian, Bo and Toma, Horia and Tuttle, Erick and Vasudevan, Vijay and Walter, Richard and Wang, Walter and Wilcox, Eric and Yoon, Doe Hyun},
	booktitle={IEEE/ACM International Symposium on Computer Architecture (ISCA)},
	pages={1--12},
	year={2017}
}

@inproceedings{2022digamma,
	title={{DiGamma}: Domain-Aware Genetic Algorithm for {HW}-Mapping Co-Optimization for {DNN} Accelerators},
	author={Kao, Sheng-Chun and Pellauer, Michael and Parashar, Angshuman and Krishna, Tushar},
	booktitle={Design, Automation \& Test in Europe Conference \& Exhibition (DATE)},
	pages={232--237},
	year={2022}
}

@inproceedings{2019timeloop,
	title={{Timeloop}: A Systematic Approach to {DNN} Accelerator Evaluation},
	author={Parashar, Angshuman and Raina, Priyanka and Shao, Yakun Sophia and Chen, Yu-Hsin and Ying, Victor A. and Mukkara, Anurag and Venkatesan, Rangharajan and Khailany, Brucek and Keckler, Stephen W. and Emer, Joel},
	booktitle={IEEE International Symposium on Performance Analysis of Systems and Software (ISPASS)},
	pages={304--315},
	year={2019}
}

@inproceedings{2019simba,
	title={{Simba}: Scaling Deep-Learning Inference with Multichip-Module-Based Architecture},
	author={Shao, Yakun Sophia and Clemons, Jason and Dreslinski, Ronald and Fowers, Jeremy and Giroux, Olivier and Hazra, Jashwanth and Hazra, Shuotao and Hegde, Kartik and Hestness, Joel and Jie, Junrui and others},
	booktitle={IEEE/ACM International Symposium on Microarchitecture (MICRO)},
	pages={14--27},
	year={2019}
}

@inproceedings{2020interstellar,
	title={{Interstellar}: Using {Halide}’s Scheduling Language to Analyze {DNN} Accelerators},
	author={Yang, Xuan and Gao, Mingyu and Liu, Qiaoyi and Setter, Jeff and Pu, Jing and Nayak, Ankita and Bell, Steven Emberton and Cao, Kaidi and Ha, Heonjae and Raina, Priyanka and Kozyrakis, Christos and Horowitz, Mark},
	booktitle={International Conference on Architectural Support for Programming Languages and Operating Systems (ASPLOS)},
	pages={369--383},
	year={2020}
}

@inproceedings{2020gamma,
	title={{GAMMA}: Automating the {HW} Mapping of {DNN} Models on Accelerators via Genetic Algorithm},
	author={Kao, Sheng-Chun and Krishna, Tushar},
	booktitle={IEEE/ACM International Conference on Computer-Aided Design (ICCAD)},
	pages={44:1--44:9},
	year={2020}
}

@inproceedings{2021hasco,
	title={{HASCO}: Towards Agile Hardware and Software Co-Design for Tensor Computation},
	author={Xiao, Qingcheng and Zheng, Size and Wu, Bingzhe and Li, Yixiao and Li, Lin and Yu, Yongpan and Xie, Yuan and Cong, Jason},
	booktitle={IEEE/ACM International Symposium on Computer Architecture (ISCA)},
	pages={1055--1068},
	year={2021}
}

@inproceedings{2023spotlight,
	title={Leveraging Domain Information for the Efficient Automated Design of Deep Learning Accelerators},
	author={Sakhuja, Chirag and Shi, Zhiyi and Lin, Calvin},
	booktitle={IEEE International Symposium on High-Performance Computer Architecture (HPCA)},
	pages={287--301},
	year={2023}
}

@article{2025senna,
	title={{SENNA}: Unified Hardware/Software Space Exploration for Parametrizable Neural Network Accelerators},
	author={Kwon, Jungyoon and Min, Hyemi and Egger, Bernhard},
	journal={ACM Transactions on Embedded Computing Systems},
	volume={24},
	number={2},
	pages={1--26},
	year={2025}
}

@article{2025memorycentric,
	title={Design Space Exploration for Scalable {DNN} Accelerators Using a Memory-Centric Analytical Model for {HW/SW} Co-Design},
	author={Huang, Wei-Chun and Tang, Chih-Wei and Chang, Kuei-Chung and Chen, Tien-Fu and Hsieh, Hsiang-Cheng and Tsai, Ming-Hsuan},
	journal={ACM Transactions on Design Automation of Electronic Systems},
	volume={30},
	number={3},
	pages={1--29},
	year={2025}
}

@article{2024tensormap,
	title={{TensorMap}: A Deep {RL}-Based Tensor Mapping Framework for Spatial Accelerators},
	author={Wang, Fuyu and Shen, Minghua and Lu, Yutong and Xiao, Nong},
	journal={IEEE Transactions on Computers},
	volume={73},
	number={8},
	pages={1899--1912},
	year={2024}
}

@inproceedings{2025factorflow,
	title={{FactorFlow}: Mapping {GEMM}s on Spatial Architectures Through Adaptive Programming and Greedy Optimization},
	author={Ronzani, Marco and Silvano, Cristina},
	booktitle={Asia and South Pacific Design Automation Conference (ASPDAC)},
	pages={706--712},
	year={2025}
}

@inproceedings{2025kapla,
	title={{KAPLA}: Scalable {NN} Accelerator Dataflow Design Space Structuring and Fast Exploring},
	author={Li, Zhiyao and Gao, Mingyu},
	booktitle={Asia and South Pacific Design Automation Conference (ASPDAC)},
	pages={8--15},
	year={2025}
}

@inproceedings{2023salsa,
	title={{SALSA}: Simulated Annealing Based Loop-Ordering Scheduler for {DNN} Accelerators},
	author={Jung, Victor J. B. and Symons, Arne and Mei, Linyan and Verhelst, Marian and Benini, Luca},
	booktitle={IEEE International Conference on Artificial Intelligence Circuits and Systems (AICAS)},
	pages={1--5},
	year={2023}
}

@inproceedings{2021mindmappings,
	title={{Mind Mappings}: Enabling Efficient Algorithm-Accelerator Mapping Space Search},
	author={Hegde, Kartik and Tsai, Po-An and Huang, Sitao and Chandra, Vikas and Parashar, Angshuman and Fletcher, Christopher W.},
	booktitle={International Conference on Architectural Support for Programming Languages and Operating Systems (ASPLOS)},
	pages={943--958},
	year={2021}
}

@inproceedings{2023dosa,
	title={{DOSA}: Differentiable Model-Based One-Loop Search for {DNN} Accelerators},
	author={Hong, Charles Jungwoo and Li, Zixuan and Shao, Yakun Sophia},
	booktitle={IEEE/ACM International Symposium on Microarchitecture (MICRO)},
	pages={209--224},
	year={2023}
}

@inproceedings{2021cosa,
	title={{CoSA}: Scheduling by Constrained Optimization for Spatial Accelerators},
	author={Huang, Qijing and Kang, Minwoo and Dinh, Grace and Norell, Thomas and Kalaiah, Aravind and Demmel, James and Wawrzynek, John and Shao, Yakun Sophia},
	booktitle={IEEE/ACM International Symposium on Computer Architecture (ISCA)},
	pages={554--566},
	year={2021}
}

@inproceedings{2021loma,
	title={{LOMA}: Fast Auto-Scheduling on {DNN} Accelerators Through Loop-Order-Based Memory Allocation},
	author={Symons, Arne and Mei, Linyan and Verhelst, Marian},
	booktitle={IEEE International Conference on Artificial Intelligence Circuits and Systems (AICAS)},
	pages={1--4},
	year={2021}
}

@inproceedings{2025trex,
	title={{T-REX}: A 68-to-567$\mu$s/Token 0.41-to-3.95 $\mu$J/Token {Transformer} Accelerator with Reduced External Memory Access and Enhanced Hardware Utilization in 16nm {FinFET}},
	author={Moon, Seunghyun and Li, Mao and Chen, Gregory K. and Knag, Phil C. and Krishnamurthy, Ram Kumar and Seok, Mingoo},
	booktitle={IEEE International Solid-State Circuits Conference (ISSCC)},
	pages={406--408},
	year={2025}
}

@inproceedings{2025edge,
	title={{EdgeDiff}: 418.4 mJ/Inference Multi-Modal Few-Step Diffusion Model Accelerator with Mixed-Precision and Reordered Group Quantization},
	author={Kim, Sangjin and Oh, Jungjun and So, Jeonggyu and Choi, Yuseon and Kim, Sangyeob and Im, Dongseok and Park, Gwangtae and Yoo, Hoi-Jun},
	booktitle={IEEE International Solid-State Circuits Conference (ISSCC)},
	pages={1--3},
	year={2025}
}

@inproceedings{2025nebula,
	title={{Nebula}: A 28nm 109.8 {TOPS}/W {3D} {PNN} Accelerator Featuring Adaptive Partition, Multi-Skipping, and Block-Wise Aggregation},
	author={Zhou, Changchun and Huang, Tianling and Ma, Yanzhe and Fu, Yuzhe and Song, Xiangjie and Qiu, Siyuan and Sun, Jiacong and Liu, Min and Li, Ge and He, Yifan and others},
	booktitle={IEEE International Solid-State Circuits Conference (ISSCC)},
	pages={412--414},
	year={2025}
}

@inproceedings{2025mega,
	title={{MEGA.mini}: A Universal Generative {AI} Processor with a New Big/Little Core Architecture for {NPU}},
	author={Han, Donghyeon and Chandrakasan, Anantha P.},
	booktitle={IEEE International Solid-State Circuits Conference (ISSCC)},
	pages={1--3},
	year={2025}
}

@article{2024multi,
	title={Multi-Objective Hardware-Mapping Co-Optimisation for Multi-{DNN} Workloads on Chiplet-Based Accelerators},
	author={Das, Abhijit and Russo, Enrico and Palesi, Maurizio},
	journal={IEEE Transactions on Computers},
	volume={73},
	number={8},
	pages={1883--1898},
	year={2024}
}

@article{2024stream,
	title={{Stream}: Design Space Exploration of Layer-Fused {DNN}s on Heterogeneous Dataflow Accelerators},
	author={Symons, Arne and Mei, Linyan and Colleman, Steven and Houshmand, Pouya and Karl, Sebastian and Verhelst, Marian},
	journal={IEEE Transactions on Computers},
	volume={73},
	number={1},
	pages={1--14},
	year={2024}
}

@inproceedings{2024feather,
	title={{Feather}: A Reconfigurable Accelerator with Data Reordering Support for Low-Cost On-Chip Dataflow Switching},
	author={Tong, Jianming and Itagi, Anirudh and Chatarasi, Prasanth and Krishna, Tushar},
	booktitle={IEEE/ACM International Symposium on Computer Architecture (ISCA)},
	pages={198--214},
	year={2024}
}

@inproceedings{2024cocco,
	title={{Cocco}: Hardware-Mapping Co-Exploration Towards Memory Capacity-Communication Optimization},
	author={Tan, Zhanhong and Zhu, Zijian and Ma, Kaisheng},
	booktitle={International Conference on Architectural Support for Programming Languages and Operating Systems (ASPLOS)},
	pages={69--84},
	year={2024}
}
	
\vspace{-5em}
\begin{IEEEbiography}[{\includegraphics[width=1in,height=1.25in,clip,keepaspectratio]{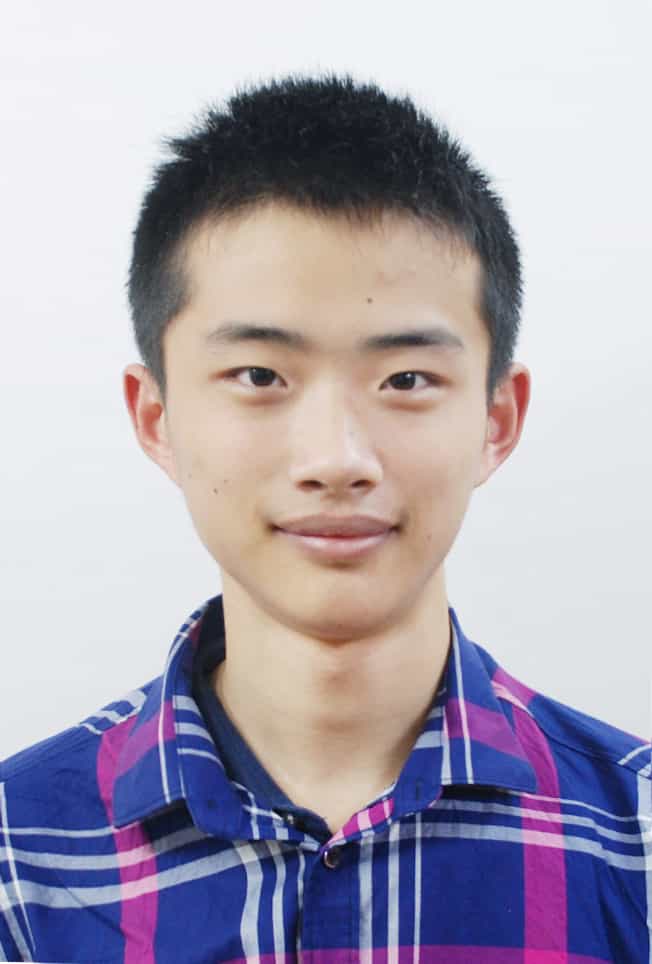}}]{Wulve Yang}
	Wulve Yang received the B.E. degree in Integrated Circuits from Huazhong University of Science and Technology, Wuhan, China, in 2023. He is currently pursuing the Ph.D. degree with the University of Chinese Academy of Sciences, Beijing, China. His research interests include neural network accelerators, hardware--software co-design, and dataflow design.
\end{IEEEbiography}
\vspace{-5em}
\begin{IEEEbiography}[{\includegraphics[width=1in,height=1.25in,clip,keepaspectratio]{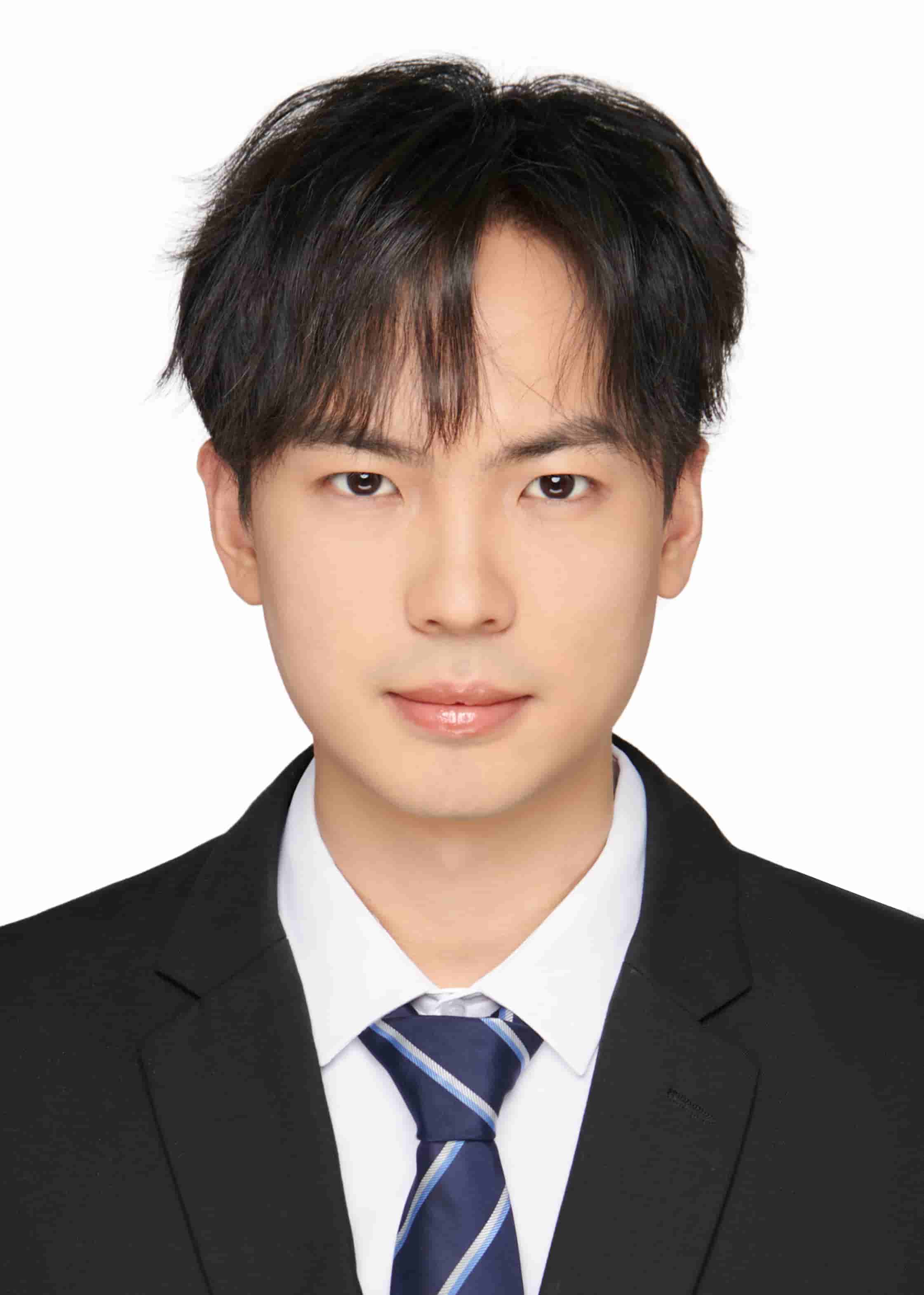}}]{Hailong Zou}
	Hailong Zou received the B.E. degree from Wuhan University, Wuhan, China, in 2023. He is currently pursuing the M.S. degree with the Institute of Microelectronics, Chinese Academy of Sciences. His research interests include keyword spotting, binary neural networks, efficient neural network design, and hardware--software co-design.
\end{IEEEbiography}
\vspace{-5em}
\begin{IEEEbiography}[{\includegraphics[width=1in,height=1.25in,clip,keepaspectratio]{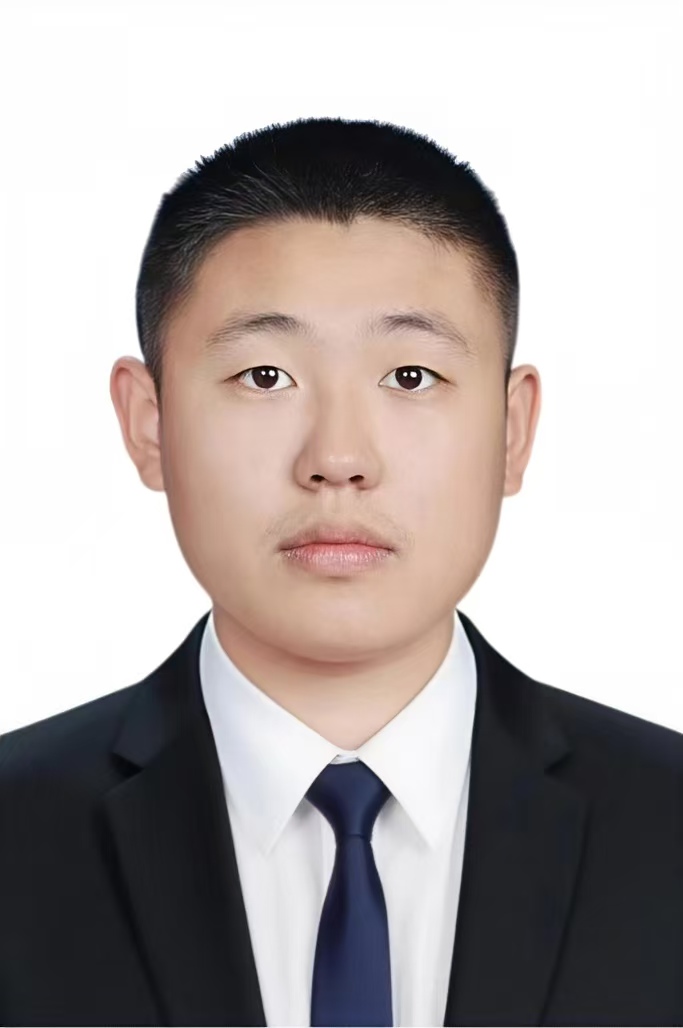}}]{Rui Zhou}
	Rui Zhou received the B.E. degree in Microelectronics Science and Technology from Tsinghua University, Beijing, China, in 2024. He is currently pursuing the Ph.D. degree with the Institute of Microelectronics, Chinese Academy of Sciences, Beijing, China. His research interests include computing-in-memory, neuromorphic chips, and mixed-precision quantization.
\end{IEEEbiography}
\vspace{-5em}
\begin{IEEEbiography}[{\includegraphics[width=1in,height=1.25in,clip,keepaspectratio]{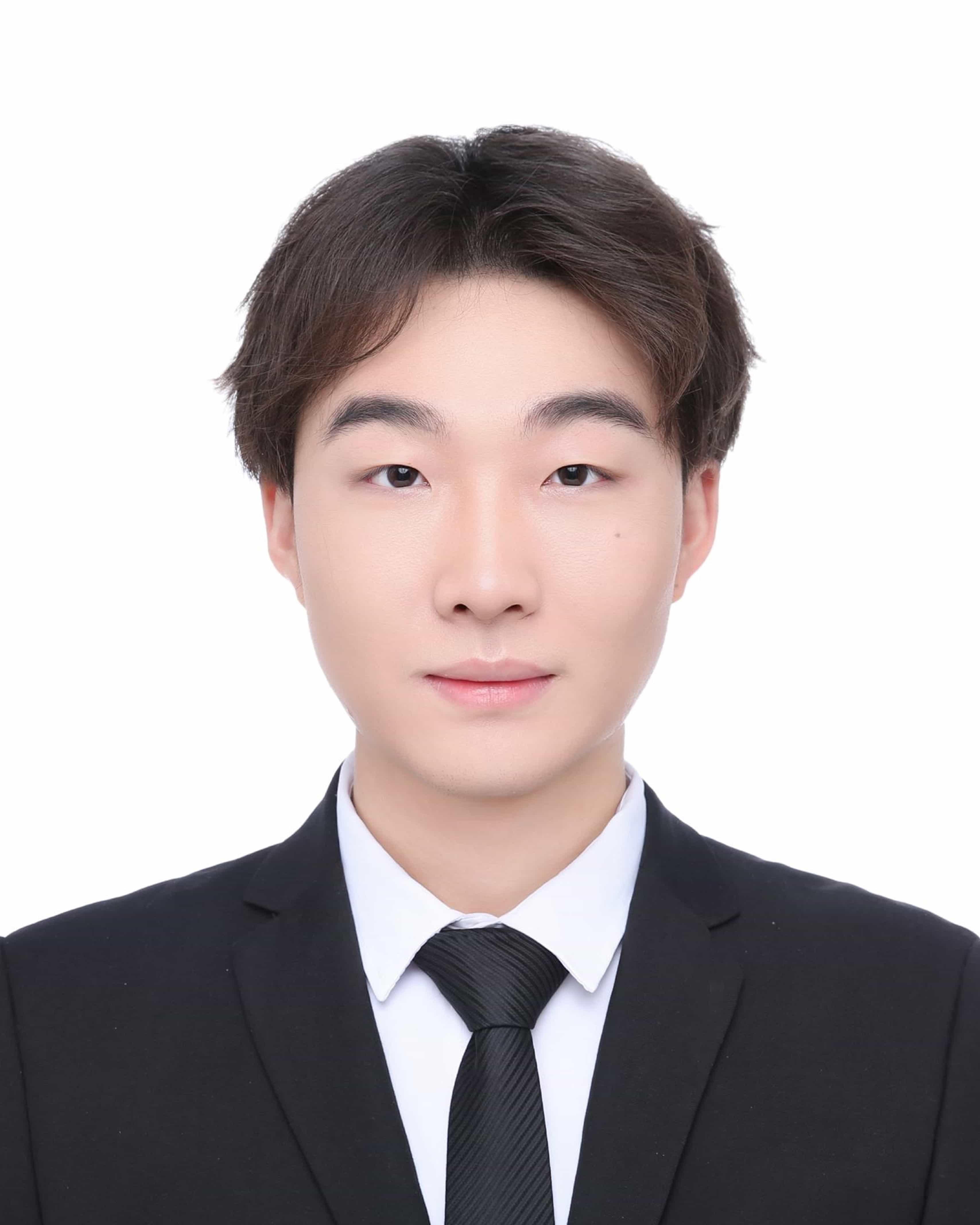}}]{Jionghao Zhang}
	Jionghao Zhang received the B.S. degree from Xidian University, Xian, China, in 2021. He is currently pursuing the Ph.D. degree with the Institute of Microelectronics of the Chinese Academy of Sciences, Beijing, China. His research interests include hardware-software co-design for neuromorphic computing, Binarized Neural Networks and Spiking Neural Networks, deep learning accelerators.
\end{IEEEbiography}

\vspace{-5em}

\begin{IEEEbiography}[{\includegraphics[width=1in,height=1.25in,clip,keepaspectratio]{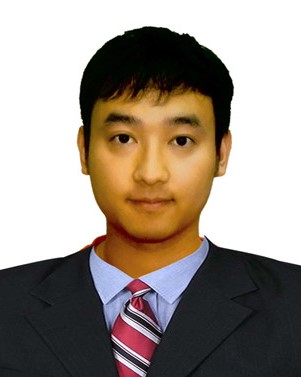}}]{Qiang Li}
	Qiang Li received the B.S. degree in Integrated Circuit Design and Integrated System from Xidian University, Xi'an, China, in 2011, the M.S. degree in Microelectronics from Peking University, Beijing, China, in 2015, and the Ph.D. degree in Electrical Engineering from Imperial College London, London, U.K., in 2020. From 2020 to 2023, he was a Senior Engineer with HiSilicon, Huawei, China. In 2023, he joined the Institute of Microelectronics, Chinese Academy of Sciences, as a Professor. His research interests include artificial intelligence and low-power design.
\end{IEEEbiography}

\vspace{-5em}

\begin{IEEEbiography}[{\includegraphics[width=1in,height=1.25in,clip,keepaspectratio]{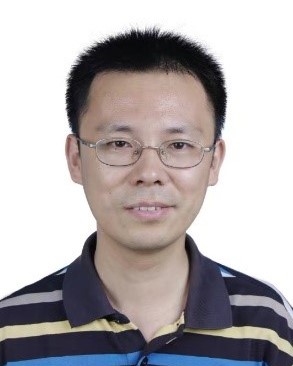}}]{Gang Li}
	Gang Li received the Ph.D. degree from the Institute
	of Microelectronics of Chinese Academy of
	Sciences, Beijing, China, in 2008. His research interests
	include energy-efficient deep learning processor
	design and system-on-chip design.
\end{IEEEbiography}

\vspace{-5em}

\begin{IEEEbiography}[{\includegraphics[width=1in,height=1.25in,clip,keepaspectratio]{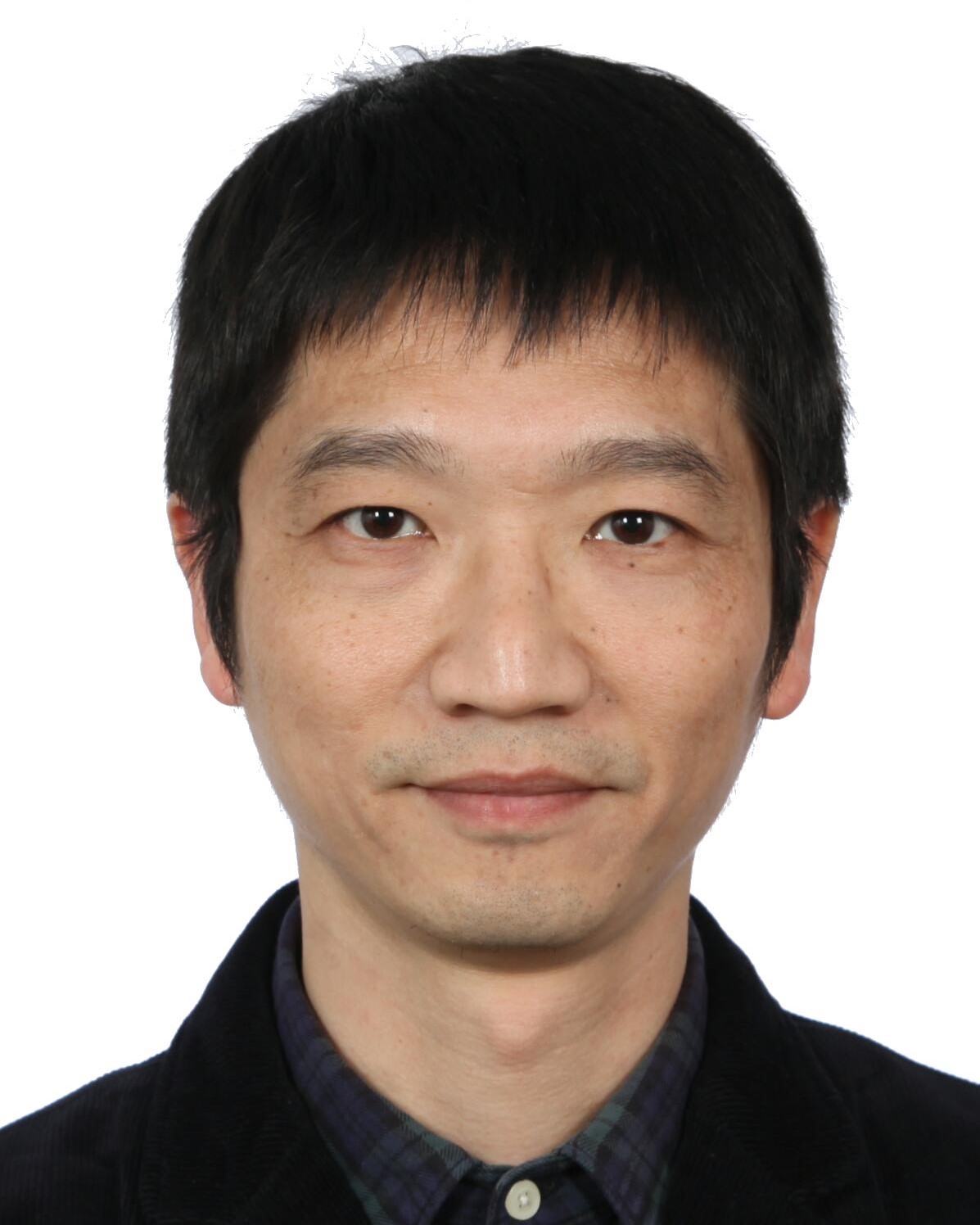}}]{Yi Zhan (Member, IEEE)}
	Yi Zhan (Member, IEEE) received the Ph.D. degree from Keio University, Yokohama, Japan, in 2012. He is currently a Professor with the Institute of Microelectronics, Chinese Academy of Sciences, Beijing, China. He is also leading a research project supported by the National Key Research and Development Program of China. His research interests include artificial intelligence, AIoT edge computing, computing-in-memory, and low-power acoustic and image microsystems. He received the Japanese Government (Monbukagakusho) MEXT Scholarship at Keio University.
\end{IEEEbiography}

\vspace{-5em}

\begin{IEEEbiography}[{\includegraphics[width=1in,height=1.25in,clip,keepaspectratio]{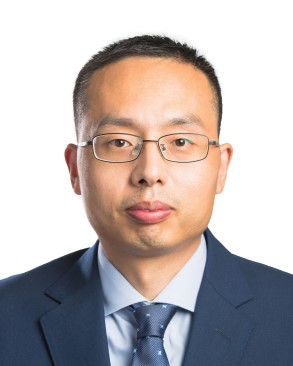}}]{Shushan Qiao (Member, IEEE)}
	Shushan Qiao (Member, IEEE) received the B.S. degree in Electronics from Hunan University, Changsha, China, in 2003, and the Ph.D. degree from the Institute of Microelectronics, Chinese Academy of Sciences, Beijing, China, in 2008. He has been a Full Professor since 2018. He is currently the Director of the Intelligent Sensing Chip and System Research Center, Institute of Microelectronics, Chinese Academy of Sciences. His research interests include computing-in-memory, artificial intelligence, ultralow-power processors, and intelligent microsystems.
\end{IEEEbiography}

\end{document}